\documentclass[a4paper,10pt,preprint,5p,number,sort&compress,twocolumn]{elsarticle}
\usepackage[utf8x]{inputenc}
\usepackage[english]{babel}
\usepackage{latexsym}
\usepackage{graphicx}
\usepackage{amsfonts}
\usepackage{amsmath}
\usepackage{amssymb}
\usepackage{braket}
\usepackage{mathtools}
\usepackage{url}

\usepackage[pdftitle={Kondo effect in a carbon nanotube with spin-orbit interaction and valley mixing: A DM-NRG study.},
pdfauthor={Mantelli, Moca, Zarand and Grifoni}]{hyperref}
\hypersetup
	{
		colorlinks=true,       
	}

\newcommand{\refEq}[1]{Eq.~(\ref{#1})}

\newcommand{\refig}[1]{Fig.~\ref{#1}}

\newcommand{\refigs}[2]{Figs.~\ref{#1} and~\ref{#2}}
\newcommand{\refsubfig}[2]{Fig.~\ref{#1}({#2})}

\newcommand{\refcite}[1]{Ref.~\cite{#1}}

\newcommand{\reapp}[1]{\ref{#1}}
 
\newcommand{\resec}[1]{Sec.~\ref{#1}}

\DeclarePairedDelimiter\abs{\lvert}{\rvert}

\newcommand{\aop}{\hat{a}}
\newcommand{\bop}{\hat{b}}
\newcommand{\cop}{\hat{c}}
\newcommand{\nop}{\hat{n}}
\newcommand{\Cop}{\hat{\mathcal{C}}}
\newcommand{\dop}{\hat{d}}
\newcommand{\fop}{\hat{f}}
\newcommand{\Hop}{\hat{H}}
\newcommand{\Pop}{\hat{\mathcal{P}}}
\newcommand{\Qop}{\hat{Q}}
\newcommand{\Top}{\hat{\mathcal{T}}}

\newcommand{\kB}{k_{\text{B}}}
\renewcommand{\vec}[1]{\textbf{#1}} 
\newcommand{\vsigma}{\boldsymbol\sigma} 
\newcommand{\vlambda}{\boldsymbol\lambda} 

\newcommand{\B}{\text{B}}
\newcommand{\bS}{\mathbf{S}}
\newcommand{\bs}{\mathbf{s}}
\newcommand{\chain}{\text{chain}}
\newcommand{\CNT}{\text{CNT}}
\newcommand{\dtxt}{\text{d}}
\newcommand{\g}{\text{g}}
\newcommand{\hc}{\text{h.c.}}
\newcommand{\HCM}{\text{HCM}}
\newcommand{\HWHM}{\text{HWHM}}
\newcommand{\orb}{\text{orb}}
\newcommand{\K}{\text{K}}
\newcommand{\KK}{\text{KK'}}
\newcommand{\lead}{\text{lead}}
\newcommand{\LL}{\text{L}}

\newcommand{\RL}{\text{R}}
\newcommand{\Rtxt}{\text{R}}
\newcommand{\s}{\text{s}}
\newcommand{\SO}{\text{SO}}
\newcommand{\SUf}{\text{SU(4)}}
\newcommand{\SUt}{\text{SU(2)}}

\newcommand{\tun}{\text{tun}}
\newcommand{\tot}{\text{tot}}
\newcommand{\Uone}{\text{U(1)}}

\begin{document}

\title{Kondo effect in a carbon nanotube with spin-orbit interaction and valley mixing: \\ A DM-NRG study.}

\author[reg]{Davide Mantelli\corref{cor1}}
\ead{davide.mantelli@physik.uni-regensburg.de}
\author[bud,ora]{C\u{a}t\u{a}lin Pa\textcommabelow{s}cu Moca}
\author[bud]{Gergely Zar\'{a}nd}
\author[reg]{Milena Grifoni}

\cortext[cor1]{Corresponding author}
\address[reg]{Institut für Theoretische Physik, Universität Regensburg, 93040 Regensburg, Germany}
\address[bud]{Department of Theoretical Physics, Institute of Physics, Budapest University of Technology and Economics, HU-1521 Budapest, Hungary}
\address[ora]{Department of Physics, University of Oradea, 410087, Oradea, Romania}

\date{\today}

\begin{abstract}
We investigate the effects of spin-orbit interaction (SOI) and valley mixing on the transport and dynamical properties of a carbon nanotube (CNT) quantum dot in the Kondo regime. As these perturbations break the pseudo-spin symmetry in the CNT spectrum but preserve time-reversal symmetry, they induce a finite splitting $\Delta$  between formerly  degenerate Kramers pairs.  Correspondingly, a crossover from the SU(4) to the SU(2)-Kondo effect occurs as the strength of these symmetry breaking parameters is varied. Clear signatures of the crossover are discussed both at the level of the spectral function as well as of the conductance. In particular, we demonstrate numerically and support with scaling arguments, that the Kondo temperature scales inversely with the splitting $\Delta$ in the crossover regime.  In presence of a finite magnetic field, time reversal symmetry is also broken.  We investigate the effects of both parallel and perpendicular fields (with respect to the tube's axis), and  discuss the conditions under which Kondo revivals may be achieved.
\end{abstract}


\maketitle

\section{Introduction}
\label{intro}
The Kondo effect \cite{Hewson1997} is a hallmark of strongly correlated electron physics. Its observation in quantum dot set-ups is ubiquitous and reveals precious information on the underlying symmetries of the quantum dot system and on the corresponding degeneracies of its spectrum. 
Specifically, electrons in carbon nanotubes (CNTs) possess a spin and a pseudo-spin degree of freedom \cite{Laird2014}, the latter originating from the presence of two inequivalent Dirac points in the underlying  graphene hexagonal lattice. In the absence of spin-orbit interaction, 
and considering  only transverse quantization,  the CNT's Hamiltonian is invariant under time-reversal and pseudo-spin reversal symmetries, 
and thus a quadruplet of degenerate levels is associated with a given longitudinal momentum. 
In this case, the four-fold degeneracy may lead to  the occurrence of the so called   $\SUf$-Kondo effect at 
low temperatures  \cite{Borda2003,ZarandSSC2003,Sasaki2004,Jarillo-Herrero2005,Choi2005}. 
In order to see this exotic Kondo resonance  it is important, however, 
that both spin and pseudo-spin quantum numbers be conserved during tunneling (or reflection), 
as a mixing of these degrees of freedom can result in a more conventional $\SUt$ Kondo effect \cite{Lim2006}. 
For CNTs devices where parts of the tube act as leads (see \refig{fig:sketch}), such a situation can be realized, and the peculiar features associated to the presence of both spin and orbital degrees of freedom can be probed in finite magnetic fields \cite{Jarillo-Herrero2005,Jarillo-HerreroPRL2005}. Recently,  $\SUf$  Kondo physics, also originating from coupled spin and orbital degrees of freedom, could be engineered in double-quantum dot based devices \cite{Keller2014}.
 
In a more realistic description of a CNT though, pseudo-spin symmetry breaking contributions, like the curvature induced spin-orbit interaction (SOI) \cite{Ando2000,Kuemmeth2008}
or valley mixing due to scattering off the boundaries \cite{Marganska2014,Izumida2015} or to disorder \cite{Kuemmeth2008}, should be included.
As a consequence, the fourfold degeneracy is broken and, for a given value of the longitudinal momentum, 
 the  spectrum of an isolated CNT quantum dot consists of two pairs of degenerate Kramers pairs, with splitting provided by the combined effects of SOI and valley mixing \cite{Jespersen2011}. In this situation, upon increasing the SOI strength or the valley mixing, a crossover from the $\SUf$-Kondo state 
 involving both Kramers pairs, to the more standard $\SUt$ Kondo regime is expected \cite{Anders2008,Galpin2010,Schmid2015}. 

Despite the considerable amount of experiments reporting Kondo behavior in CNTs,  
\cite{Jarillo-Herrero2005,Jarillo-HerreroPRL2005,Paaske2006,Quay2007,Makarovski2007,MakarovskiPRL2007,Cleuziou2013,Grove-Rasmussen2012,Schmid2015},
the combined effect of  SOI, valley mixing and the impact of applied magnetic fields on the $\SUf$ to $\SUt$ crossover have only been addressed within a field theoretical effective Keldysh action approach \cite{Schmid2015}. A numerically exact investigation of the highly intricate crossover is thus very desirable. 

In this work we study the dynamical and linear transport properties of CNT-based  Kondo quantum dots by means of the Density Matrix-Numerical Renormalization Group (DM-NRG) method~{\cite{WilsonRMP,HofstetterPRL,Anders2008,Bulla2008}. We focus on the $\SUf$ to $\SUt$ crossover induced  
by finite SOI and valley mixing, and study the influence of  magnetic fields parallel or perpendicular to the CNT's axis. At zero magnetic field, it is not possible to distinguish at the level of the spectral function or of the linear conductance among the two symmetry breaking effects. Here, what matters is the amplitude $\Delta$ of the total inter-Kramers splitting. We determine here the energy scales for the cross-over region and demonstrate that the Kondo temperature scales inversely with $\Delta$, in agreement with previous analytical predictions \cite{Schlottmann1983,Yamada1984}.

At finite fields, the behavior of the Kondo resonance is strongly influenced by the \textit{relative strength} of the SOI and valley mixing contributions, as well as by the \textit{direction} of the applied field. In fact, a major effect of the curvature induced SOI is to set as spin quantization axis the tube's axis and to lock spin and valley degrees of freedom \cite{Ando2000}. Valley mixing  instead does not act on the spin degree of freedom, but induces a rotation in valley space \cite{Jespersen2011,Marganska2014}. Thus, in parallel field the spin is still a good quantum number. Magnetic field induced Aharnov-Bohm contributions dominate over Zeeman effects at small fields, due to the large orbital moment of  the nanotubes \cite{Laird2014}, which enables one to clearly resolve the splitting \cite{Cleuziou2013} and rejoining \cite{Jespersen2011,Schmid2015} of a Kramers pair also at low fields. In perpendicular fields, in contrast, the spin is no longer a good quantum number; rather the simultaneous presence of SOI and valley mixing implies a \textit{full entanglement} of orbital and spin degrees of freedom.
In this case it is convenient to classify virtual Kondo transitions in terms of the discrete operations related to the time reversal and pseudo-spin reversal operators \cite{Schmid2015}. 
These considerations  are nicely confirmed by our simulations and reflected,  in particular, at the level of the linear conductance in the occurrence of Kondo revivals: at specific values of the magnetic field, which depend on the field direction, valley mixing strength and on the number of charges trapped in the dot, a Kondo resonance can be restored near avoided level crossings.

The paper is organized as follows. We present our model Hamiltonian for CNTs in \resec{subsec:model_ham}, and give a brief analysis of the symmetries of the system in \resec{subsec:symmetries}. Dynamical and transport properties are discussed in Secs.~\ref{sec:transport} and \ref{sec:kondo_temp}. We extend our discussion in \resec{sec:magn_field_effects} by including the effects of an applied magnetic field parallel (\resec{sec:parallel_magn}) or perpendicular (\resec{sec:perpendicular_magn}) to the CNT's axis. Our conclusions are summarized in \resec{sec:conclusion}.

\begin{figure}[tbp!]
  \centering
  \includegraphics[width=\columnwidth]{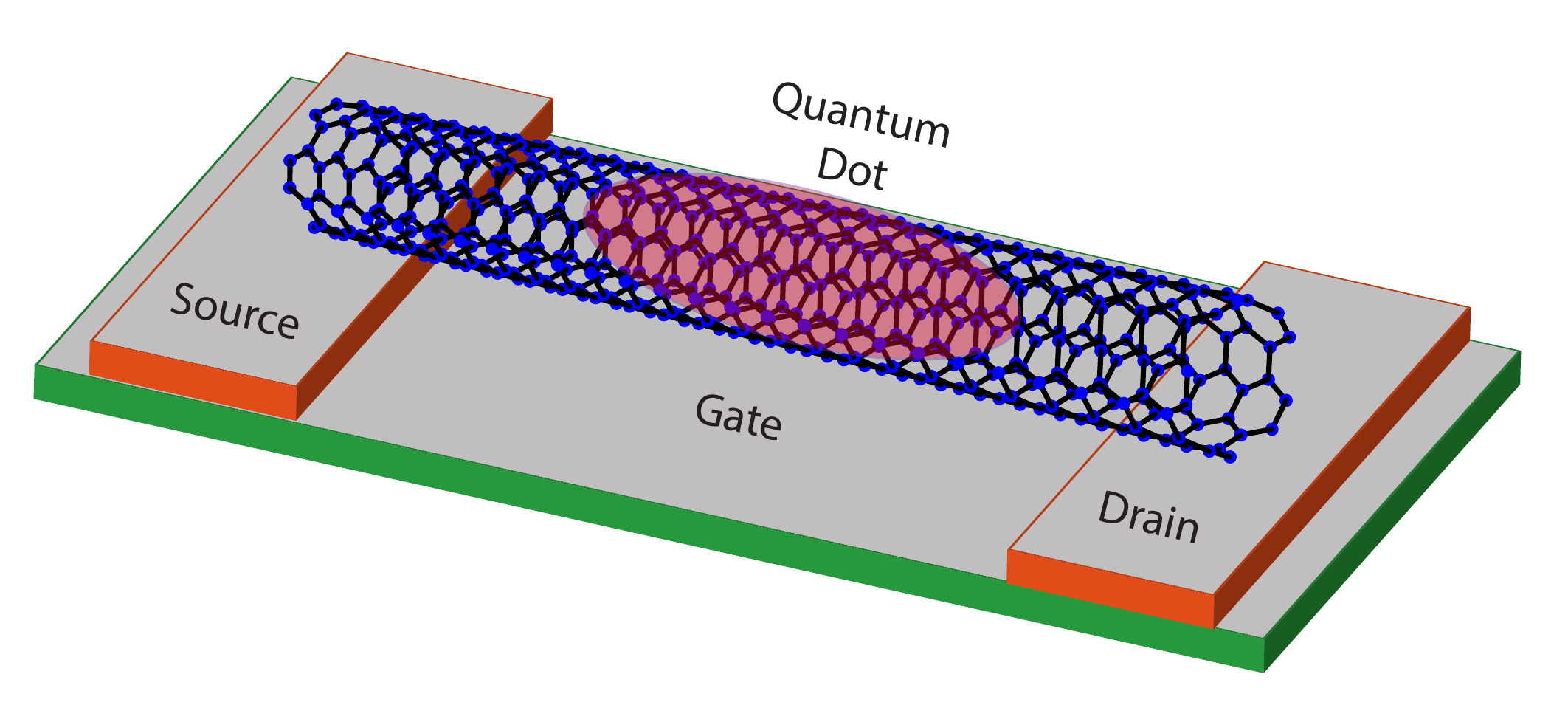}
\caption{Sketch of the setup. The carbon nanotube is coupled to two external contacts, source and drain. The quantum dot formed in the nanotube is indicated by the shaded area. The electrostatic state of the dot is capacitively controlled by a back-gate.}
\label{fig:sketch}
\end{figure}

\section{Theoretical framework}
\label{sec:theoretic_framework}

\subsection{Model Hamiltonian }
\label{subsec:model_ham}
The setup we consider consists of a CNT quantum dot coupled to two external leads (see the sketch in~\refig{fig:sketch}). We focus on a single longitudinal mode (also known as ``shell''), 
and correspondingly, describe the CNT by an extended Anderson impurity model \cite{Hewson1997,Jespersen2011}, consisting of a pair of interacting Kramers doublets.  We denote by $\varepsilon_j $ the energies  of  the four levels ($j=\{1,2,3,4\}$), and by  $\nop_j = \dop_{j}^\dagger \dop_j$ their occupation. In what follows, we shall refer to this basis as the Kramers basis (see \refsubfig{fig:CPT}{a}). Each of the four levels can accommodate one electron and, with a good approximation,  these electrons interact with each other through a strong and level-independent on-site interaction $U$. In this basis, the CNT Hamiltonian takes the form
\begin{equation}
\label{eq:H_dot}
  \Hop_{\CNT}=\sum_{j=1}^{4}\varepsilon_{j}\,\nop_j + U\sum_{j<j'}^{4} \nop_j\, \nop_{j'}\,.
\end{equation}

In the absence of the spin-orbit interaction and valley mixing,  $\Delta_{\SO}=0$ and
$\Delta_{\KK}=0$, the CNT's Hamiltonian is invariant under time-reversal and valley-reversal \cite{Laird2014}. These operations are represented 
by the two antiunitary operators $\Top$ and $\Pop$, respectively~\cite{Schmid2015}, and yield a fourfold degenerate spectrum of the CNT, $\varepsilon_j \equiv \varepsilon_{\dtxt}$. Correspondingly,  the CNT Hamiltonian is SU(4) invariant. 
In what follows, we shall label states such that $(1,2)$ and $(3,4)$ form Kramers pairs, while $(1,4)$ and $(2,3)$ are 
pairs associated with the $\Pop$ symmetry. Notice that a third unitary operator $\Cop=\Pop\Top^{-1}$  linking the remaining pairs $(1,3)$ and $(2,4)$
 can also be constructed from  $\Top$ and $\Pop$ (see \refsubfig{fig:CPT}{b}).
A finite $\Delta=\sqrt{\Delta_{\SO}^2+\Delta_{\KK}^2}$ breaks the $\Pop$ symmetry and, correspondingly, also the SU(4) symmetry (see~\ref{app:ortho} for details on how these states and the symmetry operations are constructed). 
Since time-reversal symmetry is preserved, the  on-site energies remain twofold degenerate, 
$\varepsilon_{1}=\varepsilon_{2} = \varepsilon_{\dtxt}+\Delta/2$ and $\varepsilon_{3}=\varepsilon_{4} = \varepsilon_{\dtxt}-\Delta/2$ (see \refsubfig{fig:CPT}{a}). Notice that a finite $\Delta$ plays the same role as a magnetic field on the $\mathcal{P}$ - and $\mathcal{C}$-pairs, such that conjugation relations among energy levels exist:
$\varepsilon_1(\Delta)=\varepsilon_4(-\Delta)$, and similarly for the other couples.

In a CNT, with a good approximation, each nanotube level couples 
to independent channels  in the leads, and their tunnel coupling 
can thus be described by the Hamiltonian
\begin{equation}
    \Hop_{\tun}=\sum_{j} V_j\,\sqrt{\rho_0}\int{\rm d}\varepsilon\ \aop^{\dagger}_{\varepsilon j}\dop_{j}+\hc\,.
\end{equation}
Here, instead of the original left/right operators $\cop^{\dagger}_{\varepsilon j,\LL/\RL}$ for the leads, we introduced the symmetric and antisymmetric combinations
\begin{equation*}
\label{eq:rotation_LR_lead}
	\begin{pmatrix}
	 \aop_{\varepsilon j}\\
	 \bop_{\varepsilon j}
	\end{pmatrix}=
	\begin{pmatrix}
	    \cos\gamma_{j}	&\sin\gamma_{j}\\
	    -\sin\gamma_{j}	&\cos\gamma_{j}
	\end{pmatrix}
	\begin{pmatrix}
	 \cop_{\varepsilon j,\LL}\\
	 \cop_{\varepsilon j,\RL}
	\end{pmatrix},
\end{equation*}
and the corresponding  effective tunneling amplitude $V_j\equiv \sqrt{V^{2}_{j \LL}+V^{2}_{j \RL}}$ and asymmetry parameter $\gamma_j\equiv  \arctan V_{j \RL}/V_{j \LL}$. The leads are assumed to  be non-interacting with a constant density of states per flavor $\rho\left(\omega\right)=\rho_0=1/2W$, and  a bandwidth $2W$. 
They are described by the Hamiltonian\footnote{
Quasiparticle operators
are normalized to satisfy 
$\{\aop_{\varepsilon j}, \aop_{\varepsilon' j'} \} = \delta(\varepsilon-\varepsilon')\,\delta_{jj'}$.}
\begin{equation}
 \Hop_{\lead}=\sum_{j}\int\limits_{-W}^{W} d\varepsilon\, \varepsilon\, \left(\aop^{\dagger}_{\varepsilon j} \aop_{\varepsilon j}+\bop^{\dagger}_{\varepsilon j}\bop_{\varepsilon j}\right).
\end{equation}
Notice that only the $\aop_{\varepsilon j}$ channel couples to the dot, while  
channel  $\bop_{\varepsilon j}$ remains completely decoupled in equilibrium.
The total Hamiltonian 
\begin{equation}
\Hop = \Hop_{\rm CNT}+\Hop_{\tun}+\Hop_{\lead} 
\label{eq:H_tot}
\end{equation}
captures the essential physics of our set-up and, under  equilibrium conditions,  can be solved using Wilson's NRG method \cite{WilsonRMP}. 

\subsection{Global symmetries}
\label{subsec:symmetries}

Let us now discuss the continuous symmetries of the Hamiltonian \eqref{eq:H_tot}. 
These symmetries are extremely useful, since they  allow for an efficient numerical  treatment 
of the problem. Throughout this paper, we shall focus on the simplest but physically relevant  case of 
\begin{align*}
	&V_j\equiv V,	&	&\gamma_j \equiv \gamma\;.
\end{align*}
In this case, for  $\Delta=0$,  the total SU(4)-spin operator
\begin{equation}
  \label{eq:su4_spin}
  \hat{\vec{J}}^{\SUf}=\frac{1}{2}\sum_{j,j'=1}^{4}\left(\dop^{\dagger}_{j}\vlambda_{jj'}\dop_{j'}
+   \int  {\rm d}\varepsilon\;
\aop^{\dagger}_{\varepsilon j}\vlambda_{jj'}\aop_{\varepsilon j'}\right)
\end{equation}
 commutes with the Hamiltonian~\eqref{eq:H_tot}, and the SU(4) 
symmetrical Anderson model~\cite{Choi2005} is recovered.  The $\vlambda$'s  above denote the 15 
generalized Gell Mann matrices or some other set of matrices defining the SU(4) representation, and the operators 
\eqref{eq:su4_spin} satisfy the SU(4) Lie algebra. 

Finite inter-valley scattering or spin-orbit field imply $\Delta\ne 0$, and break  the SU(4) symmetry down to  SU(2)$\otimes$ SU(2).
The latter are generated by  the usual SU(2) spin operators
\begin{equation}
  \label{eq:su2_generators}
  \hat{\vec{J}}_{\kappa}
    =\frac{1}{2}\sum_{j,j'\in\kappa}\left(\dop^{\dagger}_{j}\vsigma_{jj'}\dop_{j'}
+   \int  {\rm d}\varepsilon\;  
\aop^{\dagger}_{\varepsilon j}\vsigma_{jj'}\aop_{\varepsilon j'}\right),
\end{equation}
acting on the  two Kramers doublets $\kappa=(1,2)$ and $\kappa=(3,4)$  (see Fig.~\ref{fig:CPT}). Here,  $\vsigma=(\sigma_x,\sigma_y,\sigma_z)$ is the regular vector of the Pauli matrices.
In addition to these SU(2) symmetries, the total charge is also conserved in each Kramers 'channel'
\begin{equation}
    \Qop_{\kappa}=\frac{1}{2}\sum_{j\in\kappa}\left( \dop^{\dagger}_{j}\,\dop_{j}- 1/2 + 
     \int  {\rm d}\varepsilon\; :\aop^{\dagger}_{\varepsilon j}\aop_{\varepsilon j}:\right),
    \\
    \label{eq:U1}
\end{equation}
with $:\ldots:$ referring to normal ordering.

\begin{figure}[tbp!]
  \centering
  \includegraphics[width=\columnwidth]
  {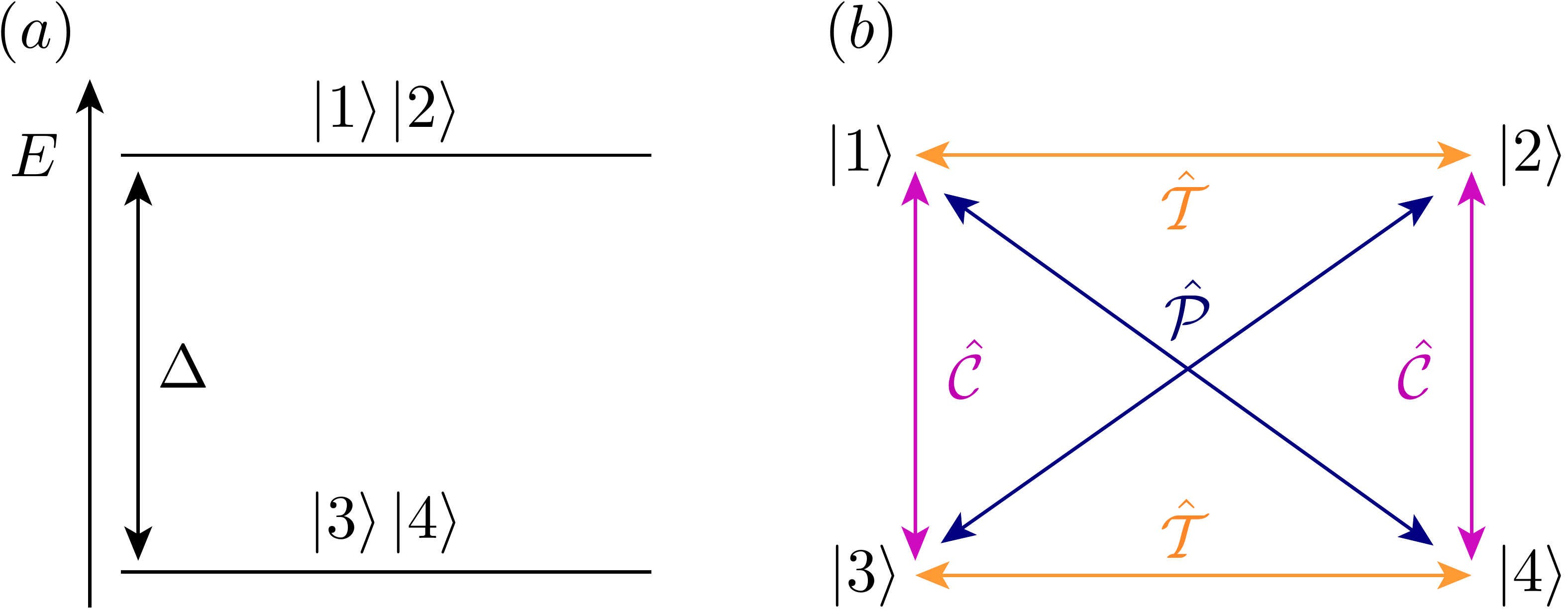}
  \caption{a) Energy level scheme involving two Kramers doublets separated by the energy $\Delta$. (b) Connection among the energy levels established by the  symmetry operations $\Top$, $\Pop$ and $\Cop$.}
  \label{fig:CPT}
\end{figure}

\section{Transport properties}
\label{sec:transport}

In this section we present the results for the spectral functions $A_j(\omega)$ of the  operators $d_j^\dagger$ and 
evaluate the conductance across the dot under equilibrium conditions. 
The linear conductance can be computed directly within the NRG and is related to the equilibrium spectral function of the 
operators $d^\dagger_j$. It reads 
\begin{equation}\label{eq:G}
 G(T)=\frac{e^2}{h} \sum_{j}\alpha_j\Gamma_{j}\int\limits^{+\infty}_{-\infty}\left(-\frac{\partial f\left(\omega, T\right)}{\partial\omega}\right)A_{j}\left(\omega\right),
\end{equation}
with $\Gamma_j=\pi V^2_j\rho_0$ the usual broadening parameter, and $\alpha_j={4\tan\gamma_j}/ \left(1+\tan\gamma_j\right)^2$ the asymmetry 
prefactor, which depends on the source and drain couplings and is in  general smaller than one.  
In \refEq{eq:G}, $f(\omega, T)=(1+\exp(\omega/T))^{-1}$ (unit $\kB=1$) is the  Fermi-Dirac distribution function and $A_j\left(\omega\right)$
denotes  the equilibrium spectral function of the $j$-th dot level,
\begin{equation}
 A_j\left(\omega\right)=-\frac{1}{\pi}{\rm Im}\bigl\{{G^{\Rtxt}_{j,j}\left(\omega\right)}\bigr\},
\end{equation} 
with $G^{\Rtxt}_{j,j'}\left(\omega\right)$ being the Fourier transform of the retarded Green's function
$
G^{\Rtxt}_{j,j'}\left(t \right) = -i\Theta (t) \langle \{\dop_j(t)\; ,   \dop^{\dagger}_{j'}(0)\}  \rangle$. To compute  $A_j\left(\omega\right)$, we used the open-access \textit{Budapest DM-NRG} code \cite{Toth2008}, that explicitly  uses the symmetries of the system. As discussed above,
for $\Delta=0$ the system exhibits SU(4) symmetry.  Therefore, by tuning the parameter $\Delta$ from 0 to some large value $\Delta \gg W$, for a singly occupied  longitudinal level, we can  follow the crossover from the the $\SUf$-Kondo fixed point to the $\SUt\otimes\SUt$ one.
In the next two subsections we shall study the manifestation of the crossover at the level of the  spectral functions and of the linear conductance, respectively. 

\subsection{Spectral functions}
\label{sec:sf_B0}

\begin{figure}[t]
  \includegraphics[angle=-90,width=\columnwidth]
  {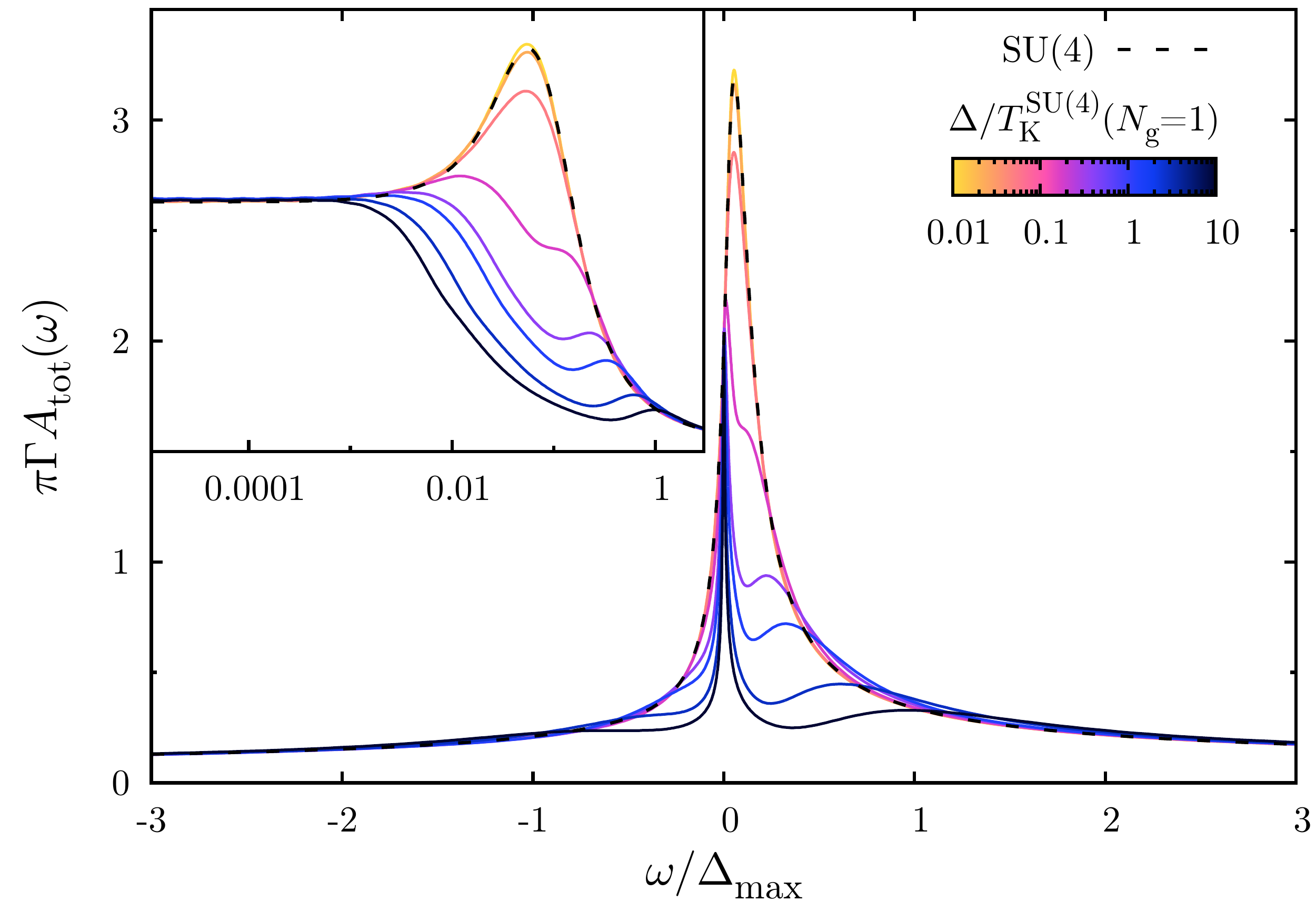}
  \caption{The total spectral function $A_{\tot}\left(\omega\right)$  as a function of frequency for different values of $\Delta$ (color scale in units of  $T^{\SUf}_{\K}=T_{\K}\left(\Delta=0\right)\simeq0.000429\Gamma$). Other parameters were fixed to $U=W$, $\varepsilon_{\dtxt}=-U/2$ ($N_{\g}=1$), $\Gamma=U/50$ and $T=0$. Inset: The evolution of the Kondo peak for different values of $\Delta$. }
  \label{fig:sf_vs_delta_ng1}
\end{figure}

\begin{figure}
  \includegraphics[angle=-90,width=\columnwidth]
  {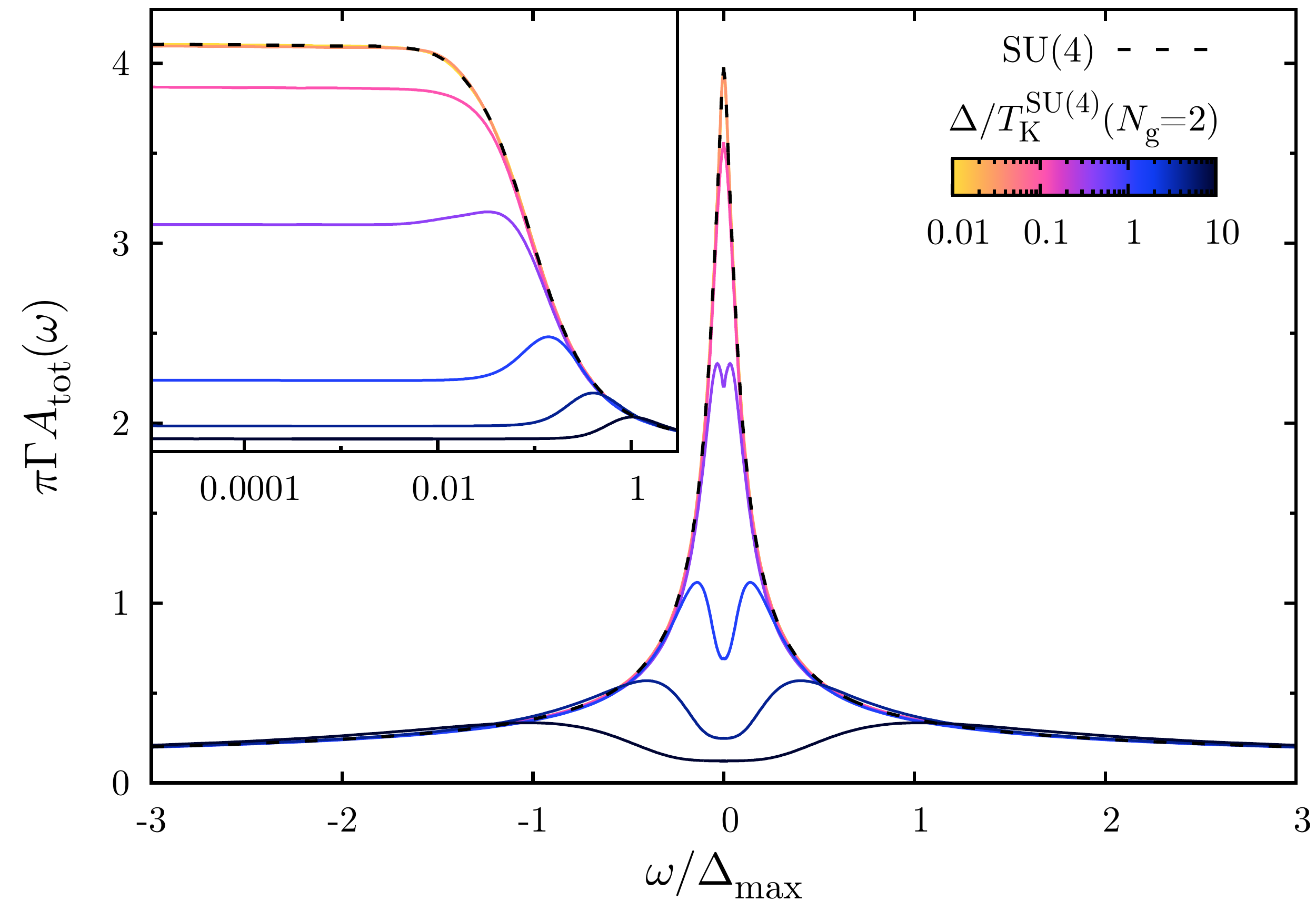}
  \caption{The $T=0$ temperature total spectral function $A_{\tot}\left(\omega\right)$  as a function of frequency for different values of $\Delta$ and at $\varepsilon_{\dtxt}=-3U/2$  ($N_{\g}=2$). We have used  $U=W$, and  $\Gamma=U/50$. (Color  scale  in units of  $T^{\SUf}_{\K}=T_{\K}\left(\Delta=0\right)\simeq0.000311\Gamma$.) Inset: Evolution of the Kondo peak on a logarithmic scale for different values of $\Delta$. }
  \label{fig:sf_vs_delta_ng2}
\end{figure}

The impact of the splitting $\Delta$ on the Kondo resonance is demonstrated in \refigs{fig:sf_vs_delta_ng1}{fig:sf_vs_delta_ng2}, where we display the  total spectral function, $A_{\tot}(\omega)=\sum_j A_j(\omega)$, for two specific values of $\varepsilon_{\dtxt}$ and a relatively  large ratio  $U/\Gamma=50$. 
For $\Delta=0$ the occupation of the CNT levels  is controlled by the 'dimensionless gate voltage', $N_{\g}=\left(-\varepsilon_{\dtxt}+U/2+\Delta\right)/\left(U+\Delta/2\right)$, taking on integer values $N_{\g}=k=1,2,3$ just in the middle of the Coulomb blockade valleys with $k$ particles on the CNT.

\refig{fig:sf_vs_delta_ng1} displays the crossover from the $\SUf$ to the $\SUt$ regime for the case $N_{\g}=1$ and several values of $\Delta$. By particle-hole symmetry,  the spectral functions for $N_{\g}=3$  are the mirror images of the  $N_{\g}=1$ spectral functions, and we do not  discuss 
them in detail. In the limit $\Delta=0$,  an $\SUf$ Kondo resonance arises
 due to quantum fluctuations of the  ground state quadruplet. As expected \cite{Hewson1997},  
 this resonance is  pinned asymmetrically to the Fermi level, $\omega=0$.
As soon as the splitting $\Delta$ becomes comparable to $T^{\SUf}_{\K}$ 
(extracted form the half width at half maximum of $A_{\tot}(\omega)$), the spectral function maximum lowers and tends to be symmetrical around the Fermi energy. At the same time, two satellite peaks emerge at approximately  $\pm\Delta$. These satellite peaks correspond to ``electron-hole'' excitations between the two Kramers pairs depicted in \refig{fig:CPT}. Notice that the value of the spectral function 
does not change at the Fermi energy as the SU(4) resonance gradually turns into a symmetrical  
SU(2) Kondo resonance. This implies that for $N_{\g}=1$ the $T=0$ temperature conductance is not suppressed by  breaking the SU(4) symmetry. However, as shown in the inset,
 the Kondo temperature is strongly reduced for $\Delta\gg T_K^{\SUf}$.

The situation is dramatically different for $N_{\g}=2$ ($\varepsilon_{\dtxt}=-3U/2$). At this value of $\varepsilon_{\dtxt}$ there are two electrons on the CNT longitudinal shell.
The Hamiltonian exhibits electron-hole symmetry, and  the total  spectral function is symmetrical for any value of $\Delta$. 
As a consequence of Friedel sum rule~\cite{Langreth66}, the value of $A_{\tot}(\omega=0)$ is twice as large as it 
was for $N_{\g}=1$. For $N_{\g}=2$, however, 
a splitting $\Delta\gg T_K^{\SUf}$ eliminates the ground state degeneracy of the isolated CNT, completely suppresses the Kondo resonance, 
and leads to the emergence of a 'pseudogap' of width $\sim\Delta$ in $A_{\tot}(\omega)$.

\subsection{Linear conductance}
\label{sec:lc_B0}

The crossover features in the spectral function  are also reflected in  transport characteristics. In \refig{fig:G_vs_vg}, we present 
the $T=0$ temperature linear conductance as a function of $N_{\g}$ for several values of the splitting $\Delta$. Similar to the spectral function, as a manifestation of the two electron $\SUf$-Kondo state, the linear conductance also acquires its maximal value at the particle-hole symmetric point $N_{\g}=2$ in the $\SUf$ symmetrical case, $\Delta=0$~\cite{Galpin2010}. This large conductance is, however, sensitive to $\Delta$ and is  quickly suppressed for $\Delta\gg T_{\K}^{\SUf}$, as a consequence of the pseudogap appearing in the spectral function. 

\begin{figure}[htbp!]
  \includegraphics[angle=-90,width=\columnwidth]
  {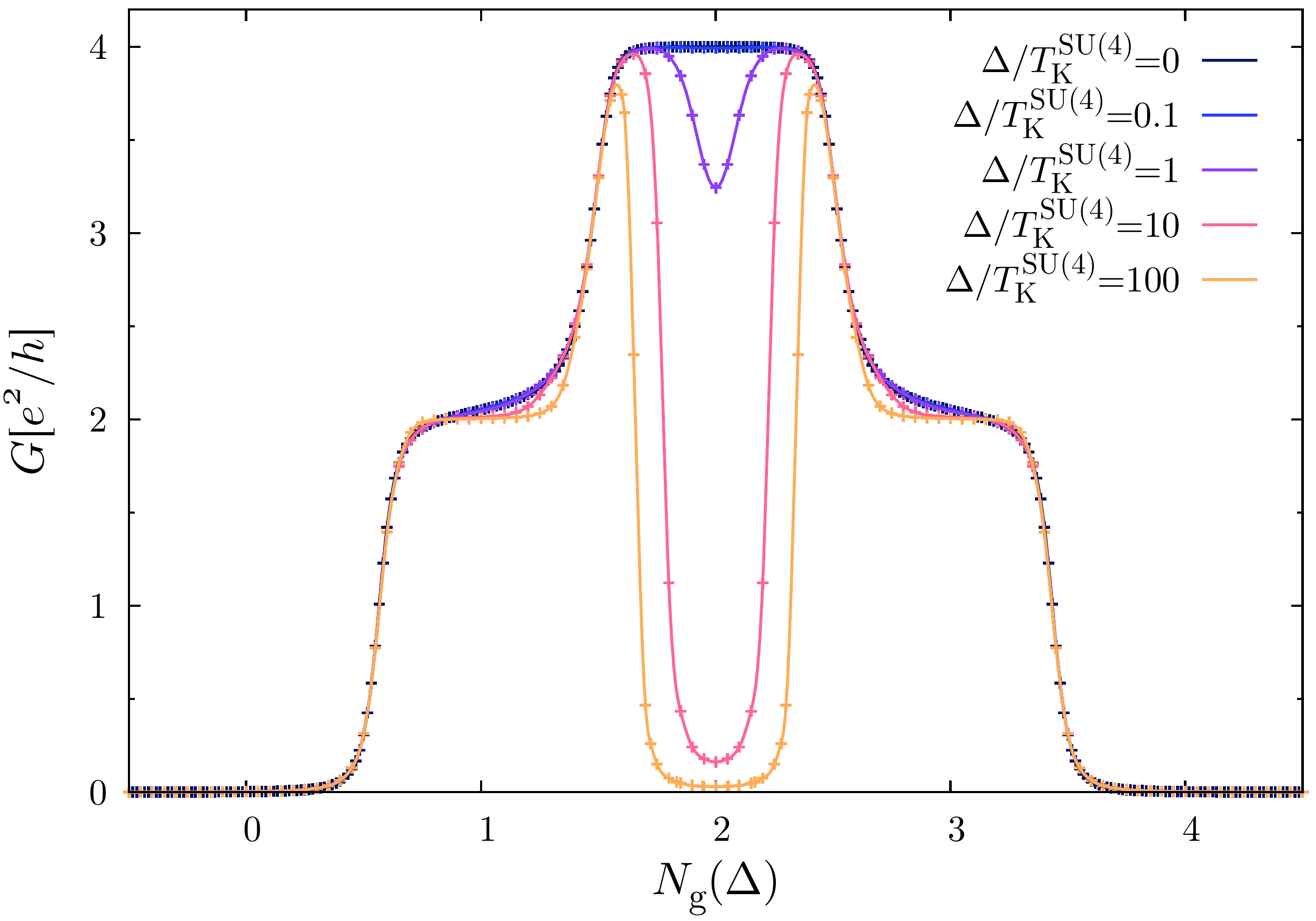}
  \caption{Linear conductance as a function of $N_{\g}$ for different values of the splitting $\Delta$ ($T^{\SUf}_{\K}=T_{\K}\left(\Delta=0,N_{\g}=1\right)=0.000429\Gamma$). The parameters are $U=W$, $\Gamma=U/50$ and $T=0$.}
  \label{fig:G_vs_vg}
\end{figure}

In contrast, for $N_{\g}=1$ and $N_{\g}=3$ a large $\Delta$ does not destroy the conductance at $T=0$ temperature, which remains close to $2e^2/h$. Notice, however, that the origin of the conductance is different in the limits $\Delta =0$ and $\Delta\gg T_{\K}^{\SUf}$. In the SU(4) limit, $\Delta\to0$, incident conduction  electrons pass through the quantum dot with probability $1/2$ in all four channels. In contrast, for $\Delta\to\infty$ the two empty levels and the corresponding channels do not conduct at all, while the other two have a perfect, unitary conductance due to the residual SU(2) Kondo effect. Since in these channels electrons pass through the dot with probability one, in this $\Delta\to\infty$ limit the 
linear conductance becomes  noiseless~\cite{Hewson1997}. 

\begin{figure}[b!]
  \includegraphics[width=\columnwidth]
  {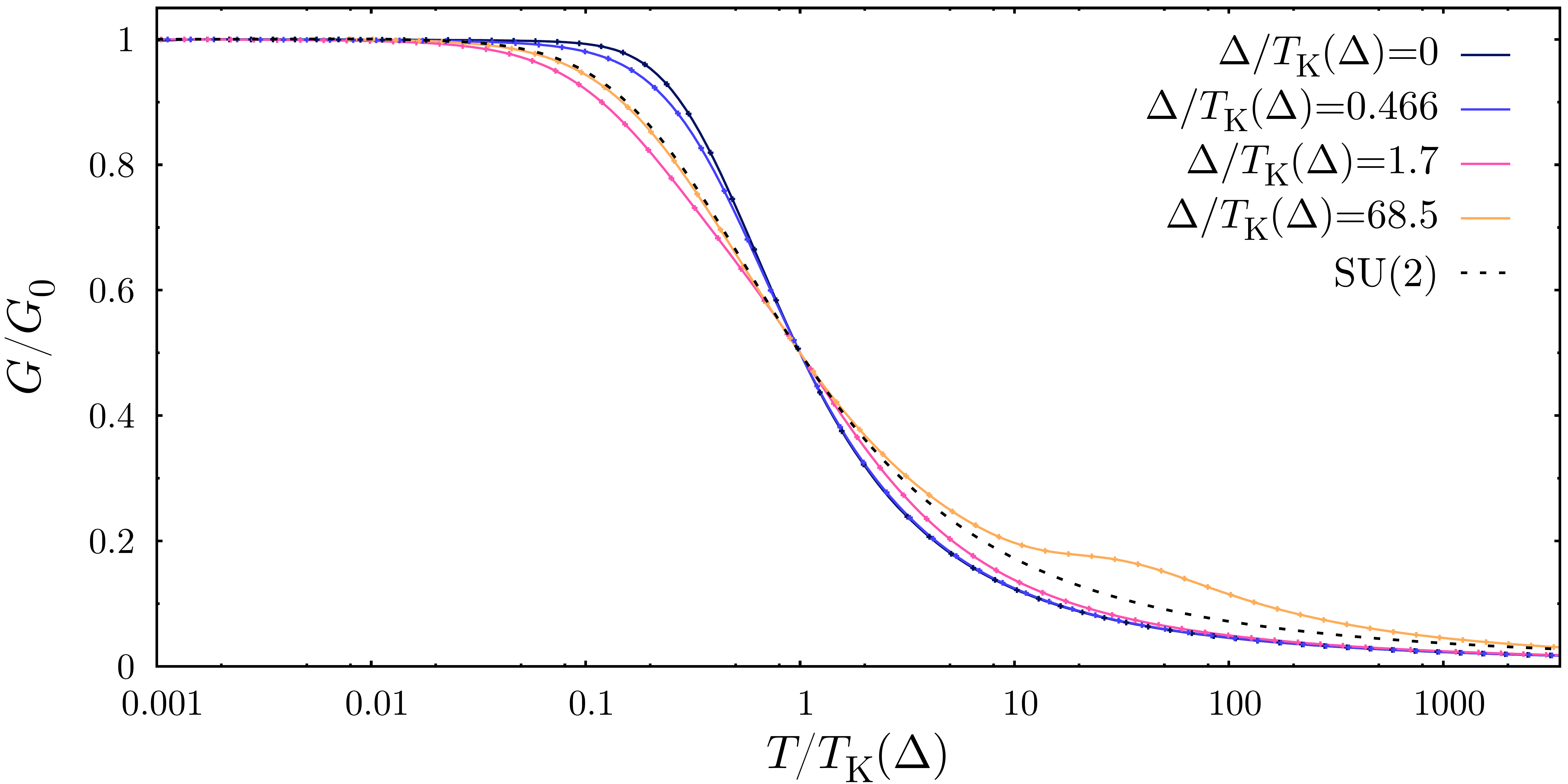}
\caption{DM-NRG result for the linear conductance as a function of $T/T_{\K}$ for different values of the splitting $\Delta$. The parameters are $U=W$, $\Gamma=U/50$, $T=0$ and $\varepsilon_{\dtxt}=-U/2$. Here $T_{\K}(\Delta)$ stands for $T^{\HCM}_{\K}(\Delta)$ defined in the main text.}
\label{fig:G_vs_TTK_delta}
\end{figure}

Another useful way to visualize the crossover between the $\SUf$  and $\SUt$ regimes is to consider the temperature dependence of the linear conductance, shown for $N_{\g}=1$ in \refig{fig:G_vs_TTK_delta}. To explore universal  scaling, we have rescaled the temperature by  the Kondo temperature $T^{\HCM}_{\K}(\Delta)$,  defined as the temperature at which the conductance is reduced to half of its $T=0$ temperature value~\cite{Hewson1997}. For $\Delta$ much smaller than  $T^{\HCM}_{\K}(\Delta)$,  conductance curves show SU(4) universality and lie on the top of each other. As soon as $\Delta$ becomes comparable with $T^{\HCM}_{\K}(\Delta)$,  however, universality is lost, the Kondo temperature lowers and a peak emerges at approximately $\Delta$. For very large values of $\Delta/T^{\HCM}_{\K}(\Delta)$, the curves become  universal again, but they now follow  a simple  $\SUt$-scaling. The two universal curves for $\Delta=0$ and $\Delta> U$ are associated with the $\SUf$ and $\SUt$ fixed points, and can be derived from the corresponding Kondo models (see \reapp{app:fermi_liquid} for further details).

\section{Kondo temperature}
\label{sec:kondo_temp}

The Kondo temperature is the most basic energy scale that characterizes the correlated Kondo  state. 
As remarked already, in the regions $N_{\g}\approx1$ and $N_{\g}\approx3$ it is 
dramatically suppressed for large Kramers pair splittings  $\Delta>T_{\K}^{\SUf}$. This suppressed SU(2) Kondo temperature, $T_{\K}^{\SUt}$, can be estimated
by carrying out  a two-step RG procedure~\cite{Eto2005}.

The Kondo temperature has first been defined in the context of a simple spin exchange Hamiltonian of the form $\Hop_{\rm int}=J\,\bS\cdot\bs$, describing the interaction of a magnetic moment $\bS$ at the origin with  the local spin density $\bs$ of the conduction electrons, 
 with $J>0$  an antiferromagnetic exchange coupling.  
Later on, it was realized that  the symmetry of the exchange Hamiltonian plays a crucial role and affects the Kondo temperature. The SU(N)-Kondo model \cite{Carmi2011}, in particular,
\footnote{The SU(N) Hamiltonian is defined in terms of an $N$-fold degenerate level as $\Hop_{\rm int} = (J/2) \; \hat X$, with $\hat X$ the exchange operator.}
yields a Kondo temperature $T_{\K}^{\rm SU(N)}\approx W \, e^{-2/N \rho_0 J}$, implying that a larger  $N$  appreciably enhances $T_{\K}$. 
This behavior can be obtained by carrying out a renormalization group analysis  for the effective 
exchange coupling (vertex function) describing scattering processes at energy $\epsilon$ . 
To leading logarithmic order, the dimensionless effective exchange coupling,  $j(\epsilon)\equiv \rho_0 J_{\rm eff}$, satisfies the RG  equation  
\begin{equation}
\label{eq:scale}
  \frac{{\rm d}j}{{\rm d} x}= \frac{N}{2} \; j^2,
\end{equation}
with  $x= \ln ( W/\epsilon)$ the scaling parameter,  and diverges logarithmically at the Kondo scale $\epsilon=T_{\K}^{\rm SU(N)}$.\footnote{Both $\epsilon$ and $W$ can be considered as scaling variables here.} 
 
Let us now focus on the Kondo regime ($\Gamma\ll U$) of a CNT quantum dot
with $N_{\g}=1$ electrons trapped inside a shell. In this regime, assuming further $\Delta\ll U$, the interaction  of the electrons with the nanotube can be described by an  exchange Hamiltonian of almost perfect SU(4) symmetry~\cite{Choi2005}. The structure of the vertex function (effective exchange amplitude), however, depends on the energy of the electrons scattered. Electrons of very high energy, $\epsilon\gg \Delta$, can induce transitions between all four levels, and experience an SU(4) exchange process. Correspondingly their scattering amplitude  satisfies  \eqref{eq:scale} with $N=4$. Electrons of energy $\epsilon\ll \Delta$, however, can only flip the states within the lower Kramers doublet, and their scattering amplitude obeys \eqref{eq:scale} with $N=2$. Matching the $N=2$ and $N=4$ solutions of \refEq{eq:scale} at $x=\ln(W/\Delta)$ we thus obtain 
\begin{equation*}
  j(\epsilon<\Delta)=\frac{1}{\ln\Bigl[{ \epsilon}\; \Delta/{\bigl(T_K^{\SUf}\bigr)^2}\Bigr]}\;.
\end{equation*}
The energy where the perturbative RG breaks down can be identified as the Kondo temperature $T_{\K}$, and is expressed as
\begin{equation}
  T_{\K}\left(\Delta\right)\propto\frac{U^2}{\Delta}\;e^{-1/j},
\label{eq:TK}
\end{equation}
predicting a $\sim 1/\Delta$ decay of the Kondo temperature for $\Delta < U$. Notice that in this expression, we replaced the bandwidth by $U$, which serves as an upper cutoff for spin exchange processes within the Anderson model. 

\begin{figure}[htb!]
  \includegraphics[angle=0,width=\columnwidth]
  {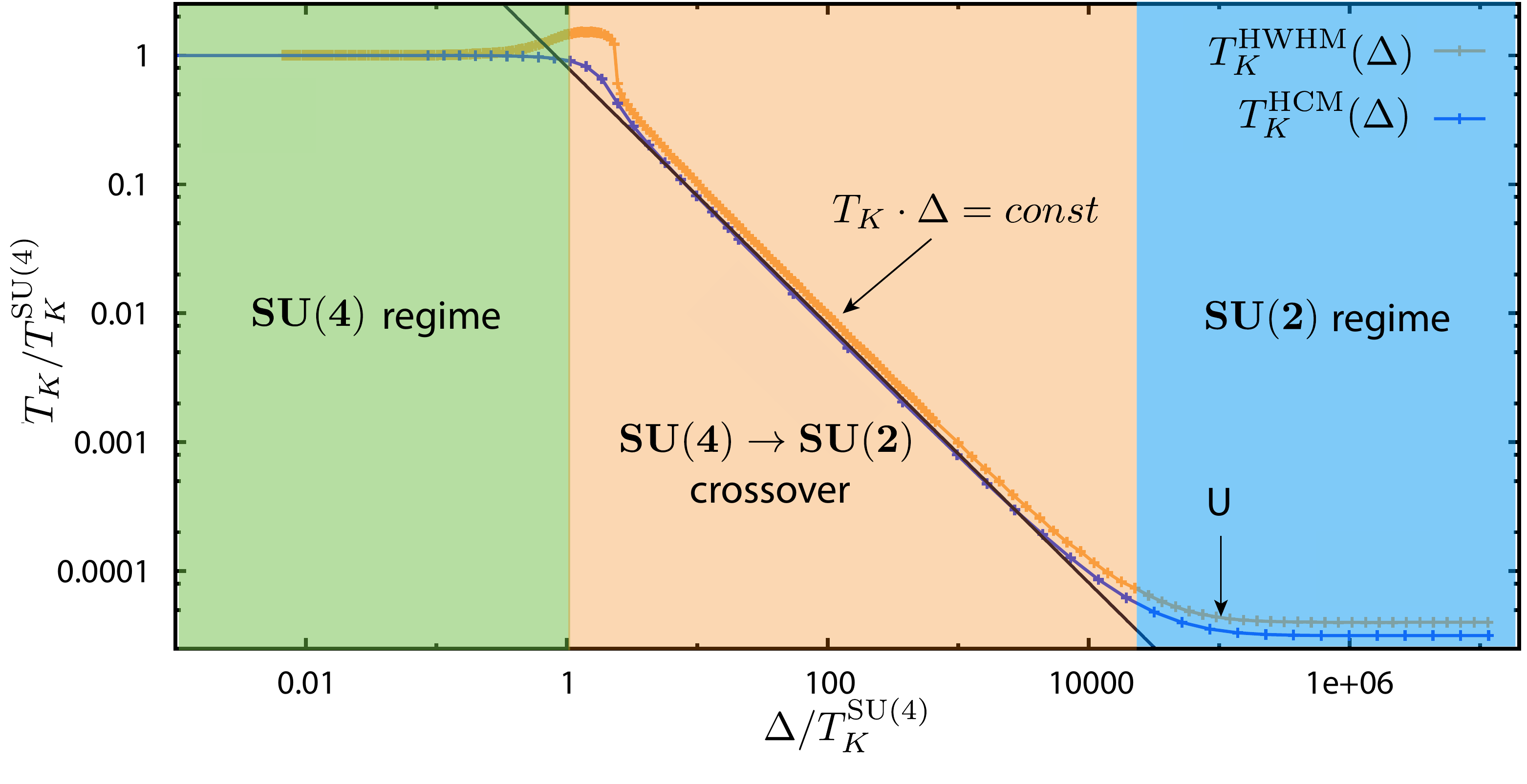}
\caption{Evolution of the Kondo temperature as a function of $\Delta$.
The Kondo temperature was extracted from the NRG data in two different ways: {\it i}) as the half width at half maximum  of the total
spectral function (HWHM),
{\it ii}) half conductance maximum (HCM). The other parameters were fixed to: $U=W$, $\varepsilon_{\dtxt}=-U/2$ ($N_{\g}=1$), $\Gamma=U/50$ and $T=0$.}
\label{fig:TK_vs_delta}
\end{figure}

As we now demonstrate, the analytical expression \eqref{eq:TK} is in good agreement with our  NRG calculations. 
The definition of  $T_{\K}$ 
is not unique, we have therefore extracted it
from the NRG data in two different ways:  in addition to 
the half conductance scale $T^{\HCM}_{\K}(\Delta)$  defined earlier, 
we also introduced the  Kondo scale $T^{\HWHM}_{\K}(\Delta)$ as the half width at half maximum of the spectral function for the $\dop_j$ operators. Both Kondo temperatures are shown in \refig{fig:TK_vs_delta}, where three regimes can be delineated: For $\Delta < T_{\K}^{\SUf}$, the system is governed by SU(4) physics, and $T_{\K}$   agrees with the SU(4) Kondo temperature. This is followed by a 'crossover region', $T_{\K}^{\SUf}<\Delta< U$, where an SU(4)$\to$ SU(2)  crossover takes place as a function of temperature or energy, but below $\Delta$ an SU(2) Kondo state emerges  with  a suppressed 
Kondo temperature given by \eqref{eq:TK}. Finally, 
for $\Delta>U$ only one Kramers doublet remains active, and an SU(2) Kondo behavior appears at all scales 
with a $T_{\K}$ independent of $\Delta$.

\section{Transport in a finite magnetic field}

\label{sec:magn_field_effects}
A finite magnetic field turns out to be crucial to disentangle the effects of SOI and valley mixing. As shown below, both the spectral function and the linear conductance display qualitatively different features depending on the direction of the applied magnetic field.

In finite magnetic field the $N_{\g}=1$ (center of the first Coulomb valley) condition becomes $\varepsilon_{\dtxt}=-U/2-(\varepsilon_4(B_{\parallel})+\varepsilon_3(B_{\parallel}))/2$, whereas $N_{\g}=2$ (particle-hole symmetric point) remains $\varepsilon_{\dtxt}=-3U/2$.

\subsection{Parallel magnetic field}
\label{sec:parallel_magn}

\begin{figure}[b!]
  \includegraphics[width=\columnwidth]
  {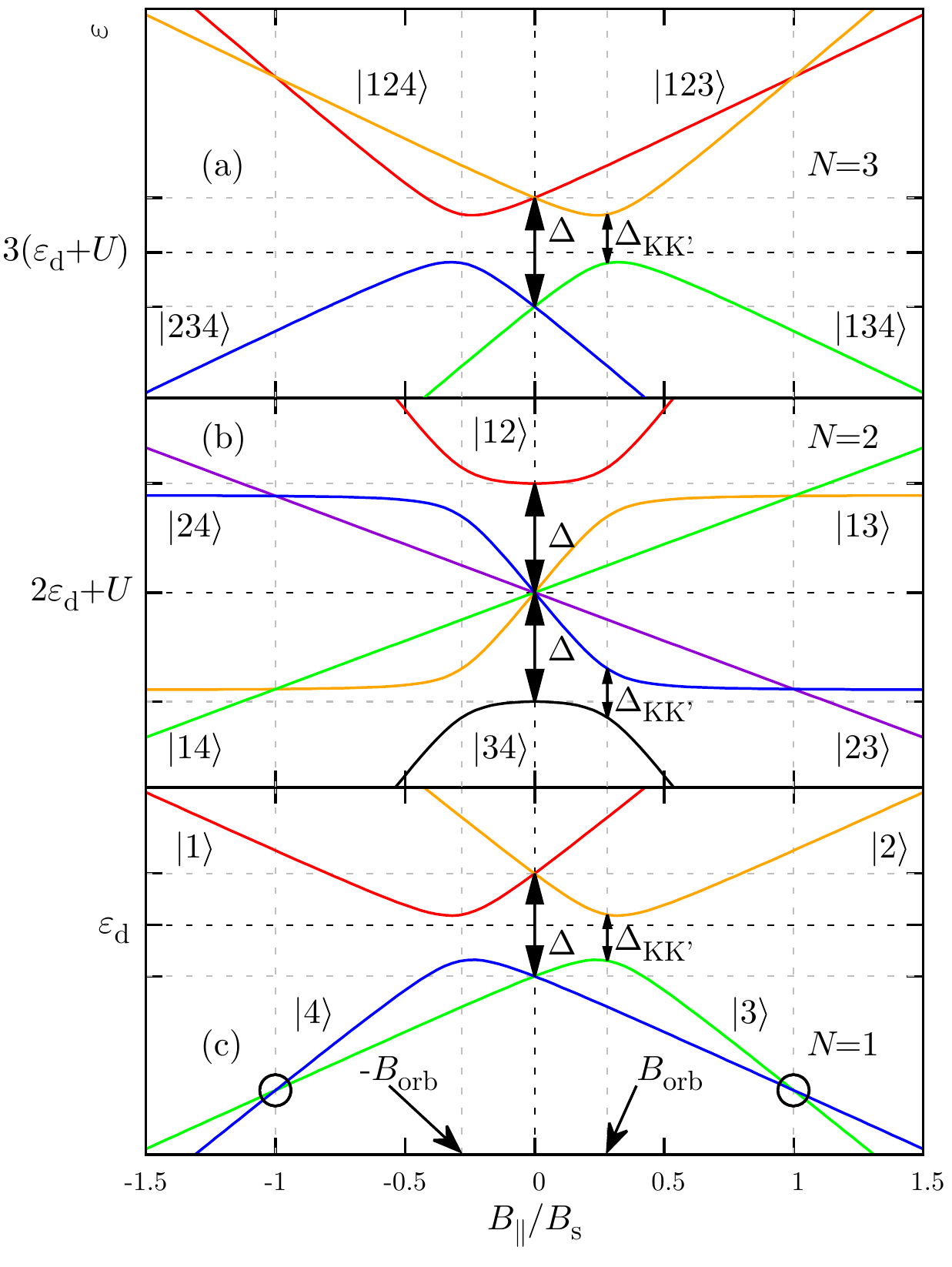}
  \caption{Parallel magnetic field dependence of the energy levels for (a) $N=3$, (b) $N=2$ and (c) $N=1$ Hilbert space sectors. Black circles indicate ground-state levels crossing. The parameters are $U=W$, $\Gamma=0.02W$,  $\varepsilon_{\dtxt}=-U/2$, $g_{\s}=2$, $g_{\orb}=1.83g_{\s}$ and $\Delta_{\KK}=\Delta_{\SO}/2=5\Gamma$. They are chosen such that the inequality \eqref{eq:Bs_condition} is fulfilled.}
\label{fig:energy_spectrum}
\end{figure}
\begin{figure}[htbp!]
  \includegraphics[angle=-90,width=\columnwidth]
  {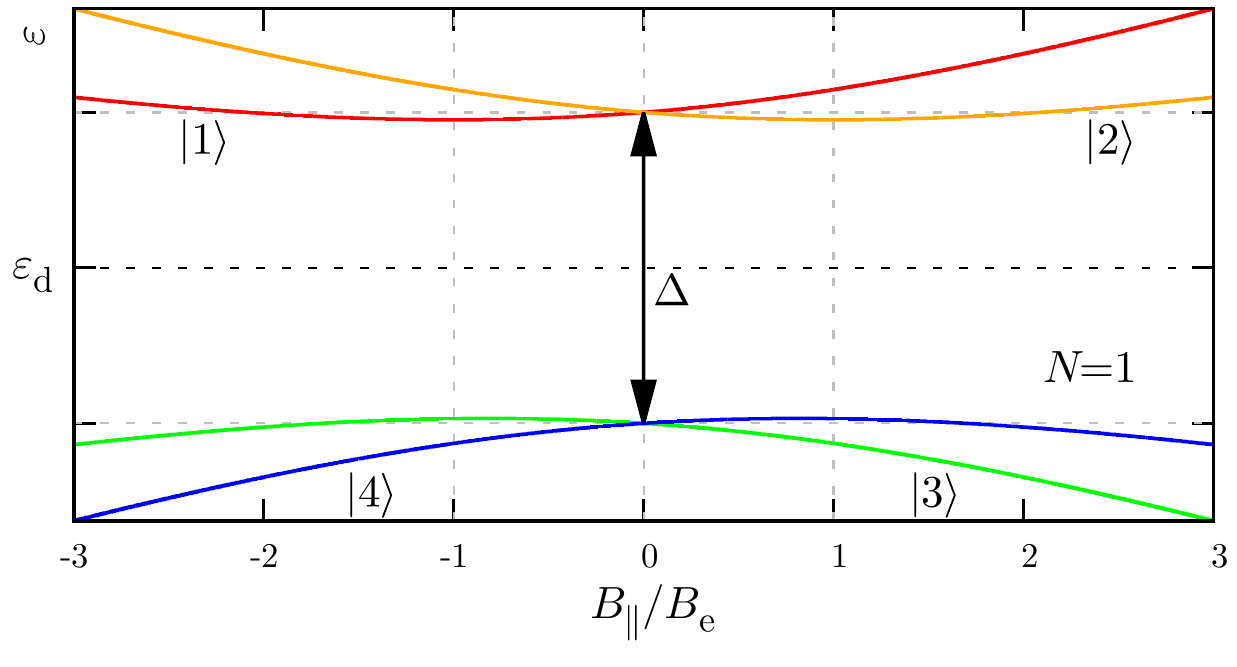}
  \caption{Evolution of the single particle energy levels in parallel magnetic field. The parameters are $\Gamma=0.02W$, $U=W$, $\varepsilon_{\dtxt}=-U/2$, $\Delta_{\SO}=5\Gamma$, $g_{\s}=2$, $g_{\orb}=1.83g_{\s}$ and $\Delta_{\KK}=10\Delta_{\SO}\sqrt{4g_{\orb}^{2}/g_{\s}^{2}-1}$. For this choice of parameters the inequality \eqref{eq:Bs_condition} is not satisfied and no level crossing is possible at finite magnetic field. Here, $B_{\text{e}}>0$ corresponds to the value of $B_{\parallel}$ such that the state $\Ket{2}$ shows a minimum.}
\label{fig:energy_spectrum_2}
\end{figure}

\paragraph{Spectrum} The energy levels of the  CNT Hamiltonian,~\refEq{eq:H_dot}, easily follow from the discussion in \reapp{subsec:specttrum_par_field}. They read
\begin{subequations}
\label{eq:B_parallel_es}
 \begin{align}
    \varepsilon_{1,4}&=\varepsilon_{\dtxt}+\frac{1}{2}g_sB_{\parallel}\pm\frac{1}{2}\sqrt{\Delta_{\KK}^2+\left(\Delta_{\SO}+2g_{\orb}B_{\parallel}\right)^2},\\
    \varepsilon_{2,3}&=\varepsilon_{\dtxt}-\frac{1}{2}g_{\s}B_{\parallel}\pm\frac{1}{2}\sqrt{\Delta_{\KK}^2+\left(\Delta_{\SO}-2g_{\orb}B_{\parallel}\right)^2},
  \end{align}
\end{subequations}
where $g_{\s}$($g_{\orb}$) is the Landè spin(orbital) $g$-factor and $B_{\parallel}$ the amplitude of the parallel magnetic field.

The resulting spectrum for occupancies with $N=1,2,3$ of the CNT is shown in \refig{fig:energy_spectrum}. It is interesting to notice the crossing of the states $\Ket{3}$ and $\Ket{4}$ in the sector $N=1$ (circles in \refsubfig{fig:energy_spectrum}{c}). It can occur if 
\begin{equation}
\label{eq:Bs_condition}
 \frac{\abs{\Delta_{\KK}}}{\abs{\Delta_{\SO}}}<\sqrt{\frac{4g_{\orb}^{2}}{g_{\s}^{2}}-1},
\end{equation}
at a magnetic field value given by
\begin{equation}
  \label{eq:Bs}
 B_{\s}=\sqrt{\frac{\Delta_{\SO}^2}{g_{\s}^2}-\frac{\Delta_{\KK}^2}{4g_{\orb}^2-g_{\s}^2}}.
\end{equation}
Thus, the ground states in the one-particle sector of the Hilbert space switch from $\Ket{4}$ to $\Ket{3}$ for $B_{\parallel}>B_{\s}$. Moreover, an avoided crossing occurs at $B_{\parallel}=B_{\orb}=\Delta_{\SO}/(2g_{\orb})$. If~\refEq{eq:Bs_condition} is not satisfied, the crossing of the two states in the $N=1$ sector cannot be achieved, as it is shown in \refig{fig:energy_spectrum_2}.
\begin{figure}[htbp!]
  \includegraphics[width=\columnwidth]
  {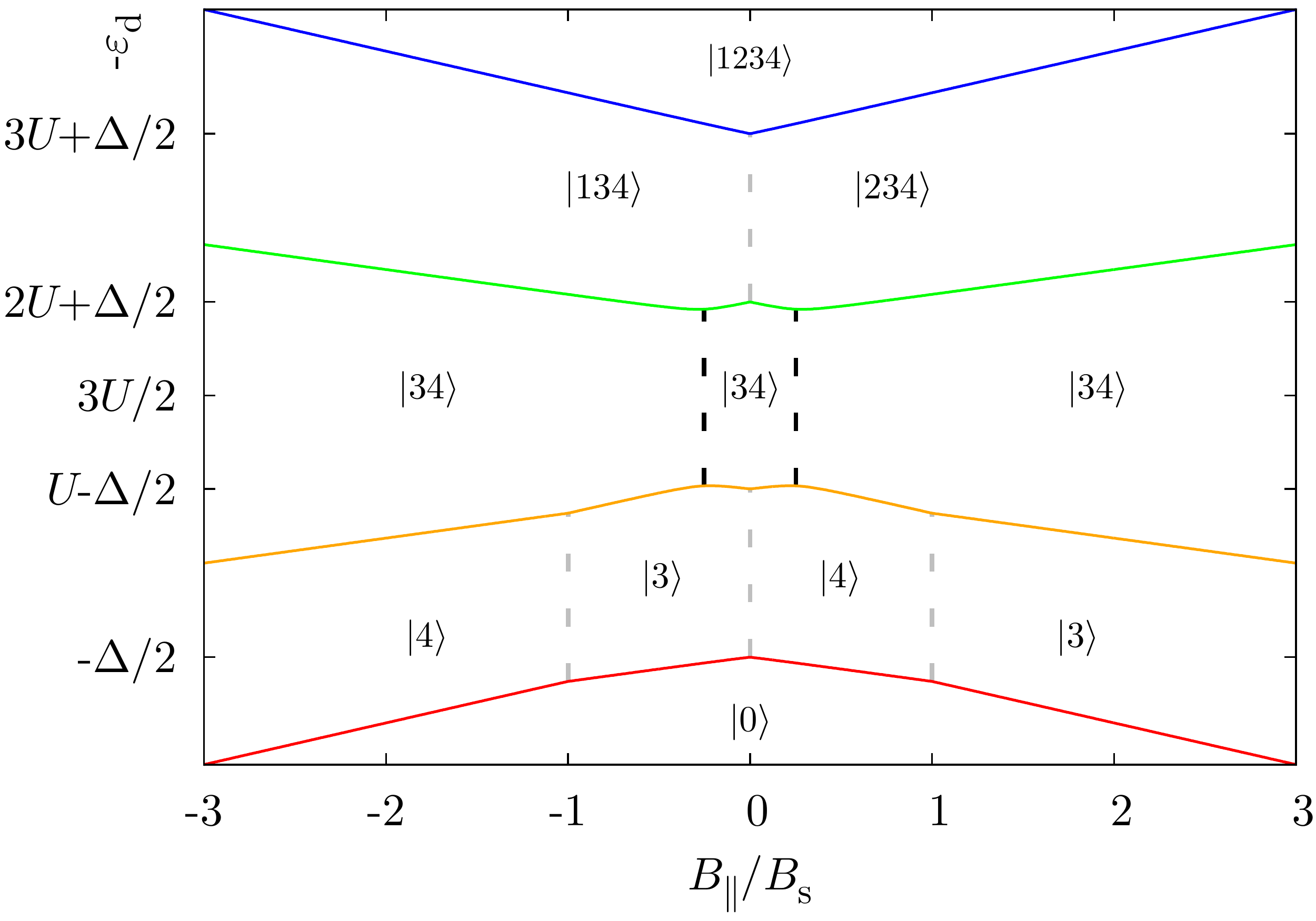}
  \caption{Ground state configuration as a function of the energy $\varepsilon_{\dtxt}$ and the applied parallel magnetic field. Grey dashed lines indicate degeneracies. The black dashed line indicates $B_{\parallel}=B_{\orb}$. Parameters are, as in \refig{fig:energy_spectrum}, $U=W$, $\Gamma=0.02W$,  $\varepsilon_{\dtxt}=-U/2$, $g_{\s}=2$, $g_{\orb}=1.83g_{\s}$ and $\Delta_{\KK}=\Delta_{\SO}/2=5\Gamma$.}
\label{fig:boundaries_states}
\end{figure}

Focusing on the $N=2$ sector of the Hilbert space (see \refsubfig{fig:energy_spectrum}{b}), one notices that again an avoided crossing occurs at $B_{\orb}=\Delta_{\SO}/2g_{\orb}$, which becomes an exact crossing for $\Delta_{\KK}=0$. In this special case the two states $\Ket{34}$ and $\Ket{24}$ are degenerate. Finally, no ground state crossing is observed for $N=3$ (see \refsubfig{fig:energy_spectrum}{a}).

Using these considerations, we can build up the ground state configuration of the system as a function of the applied parallel magnetic field and of the orbital energy $\varepsilon_{\dtxt}$ (see \refig{fig:boundaries_states}). By inspecting where a crossing occurs (dashed lines in \refig{fig:boundaries_states}), it is possible to capture the so-called ``Kondo revivals''~\cite{Galpin2010}, where the Kondo effect is restored for some specific values of the applied magnetic field. As~\refig{fig:boundaries_states} reveals, revivals are expected around $B_{\parallel}=B_{\orb}$ for $\Delta_{\KK}\simeq0$ in the $N=2$ valley, and around $B_{\parallel}=B_{\s}$, if the inequality \eqref{eq:Bs_condition} is fulfilled, in the $N=1$ valley.

These considerations are reflected in the behavior of the spectral function and of the linear conductance.

\begin{figure}[b!]
  \includegraphics[angle=-90,width=\columnwidth]
  {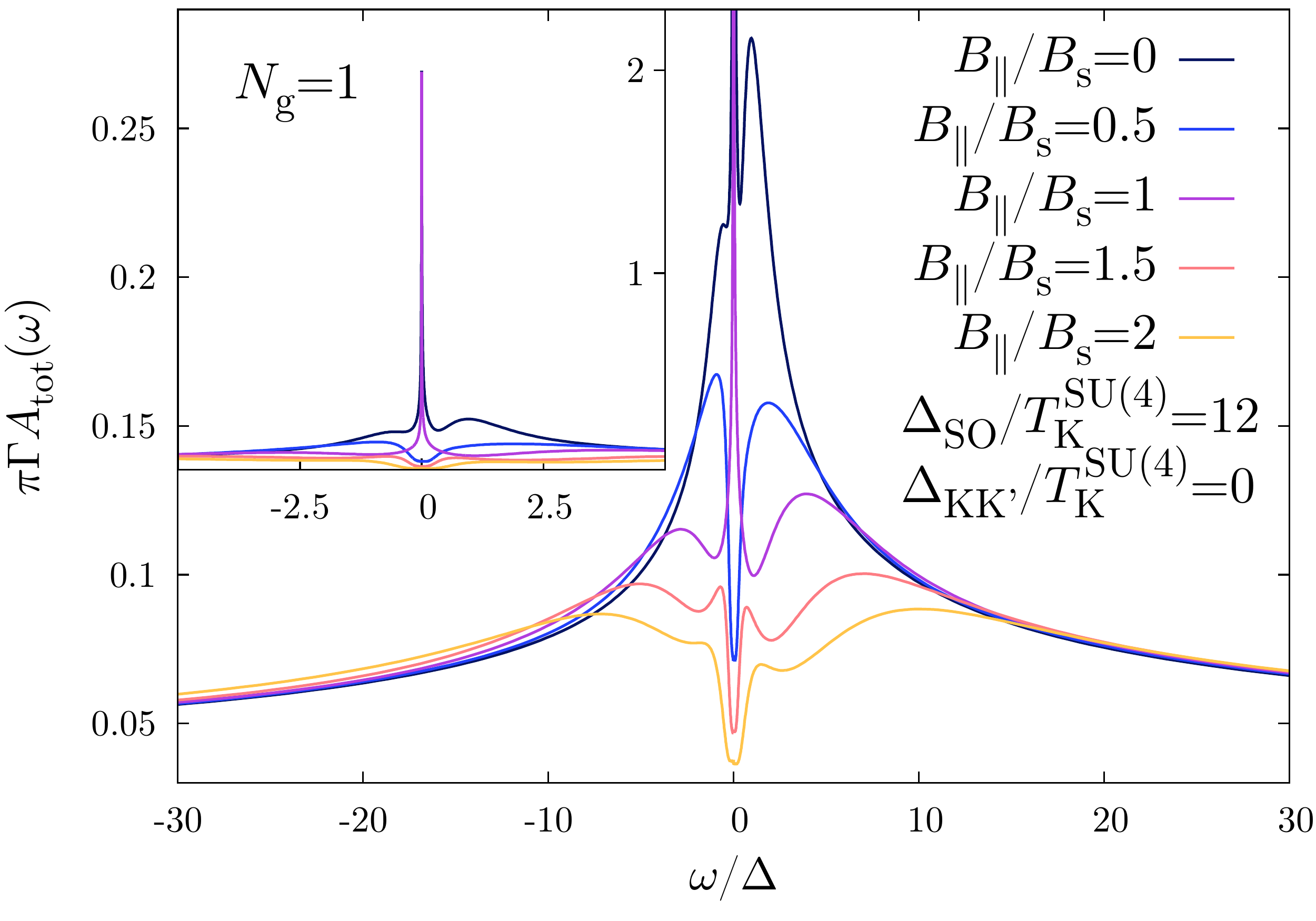}
  \caption{Close-up image of the spectral function as a function of the frequency for $\Delta_{\KK}=0$ and several values of the parallel magnetic field in the middle of the Coulomb valley  with $N_{\g}=1$ at $T=0$. Inset: Full spectral function. The unitary limit is reached for $B_{\parallel}=0$ while the full Kondo revival is obtained at $B_{\parallel}=B_{\s}$.}
\label{fig:sf_vs_omegaoverdelta}
\end{figure}

\paragraph{Spectral function} The spectral function is shown for several values of the parallel magnetic field in \refig{fig:sf_vs_omegaoverdelta}. In the absence of the magnetic field (black curve) it shows a Kondo peak at the Fermi level and two satellite peaks located at $\omega\simeq\pm\Delta=\pm(\varepsilon_{1(2)}(0)-\varepsilon_{3(4)}(0))$. As we switch on the magnetic field, the Kondo peak lowers and then splits into two resonances located at $\omega\simeq\pm(\varepsilon_3(B_{\parallel})-\varepsilon_4(B_{\parallel}))$, due to processes representing quantum fluctuations within the lowest Kramers pair. Increasing the magnetic field, the peaks merge again into one at $B_{\parallel}=B_{\s}$ (violet curve) and the spectral function recovers the unitary value at $\omega=0$ (see inset of \refig{fig:sf_vs_omegaoverdelta}). For larger values of $B_{\parallel}$ the Kondo peak splits again. Regarding the satellite peaks, they lower, broaden and shift towards higher energies. We find that at $B_{\parallel}=B_{\s}$ the leftmost one is due to a resonance in the third and fourth components ($A_{3}(\omega)=A_4(\omega)$, $\varepsilon_3=\varepsilon_4$ for $B_{\parallel}=B_{\s}$) of the spectral function. On the other hand, the rightmost peak has an internal substructure being the sum of a maximum in the first, $A_1(\omega)$, and second, $A_2(\omega)$, components. Increasing furthermore the magnetic field ($B_{\parallel}>B_{\s}$) the satellites peaks lower and broaden, the leftmost being determined essentially by the component corresponding to the ground state, $A_3(\varepsilon)$. Despite the fact that the energy separation, $\varepsilon_{2}(B_{\parallel})-\varepsilon_{1}(B_{\parallel})$, of the excited states $\Ket{1}$ and $\Ket{2}$ is monotonically increasing, the total spectral function does not show a split of the corresponding resonances in this range of investigated parallel magnetic field values ($0\leq B_{\parallel}\leq 2B_{\s}$ with $B_{\s}\simeq90T_{\K}(\Delta)$). Our results on the impact of a parallel magnetic field on the spectral function  of a CNT are consistent with the NRG analysis performed in \refcite{Lim2006} when $\Delta_{\SO}=\Delta_{\KK}=0$.

\begin{figure}[!htbp]
  \includegraphics[angle=-90,width=\columnwidth]
  {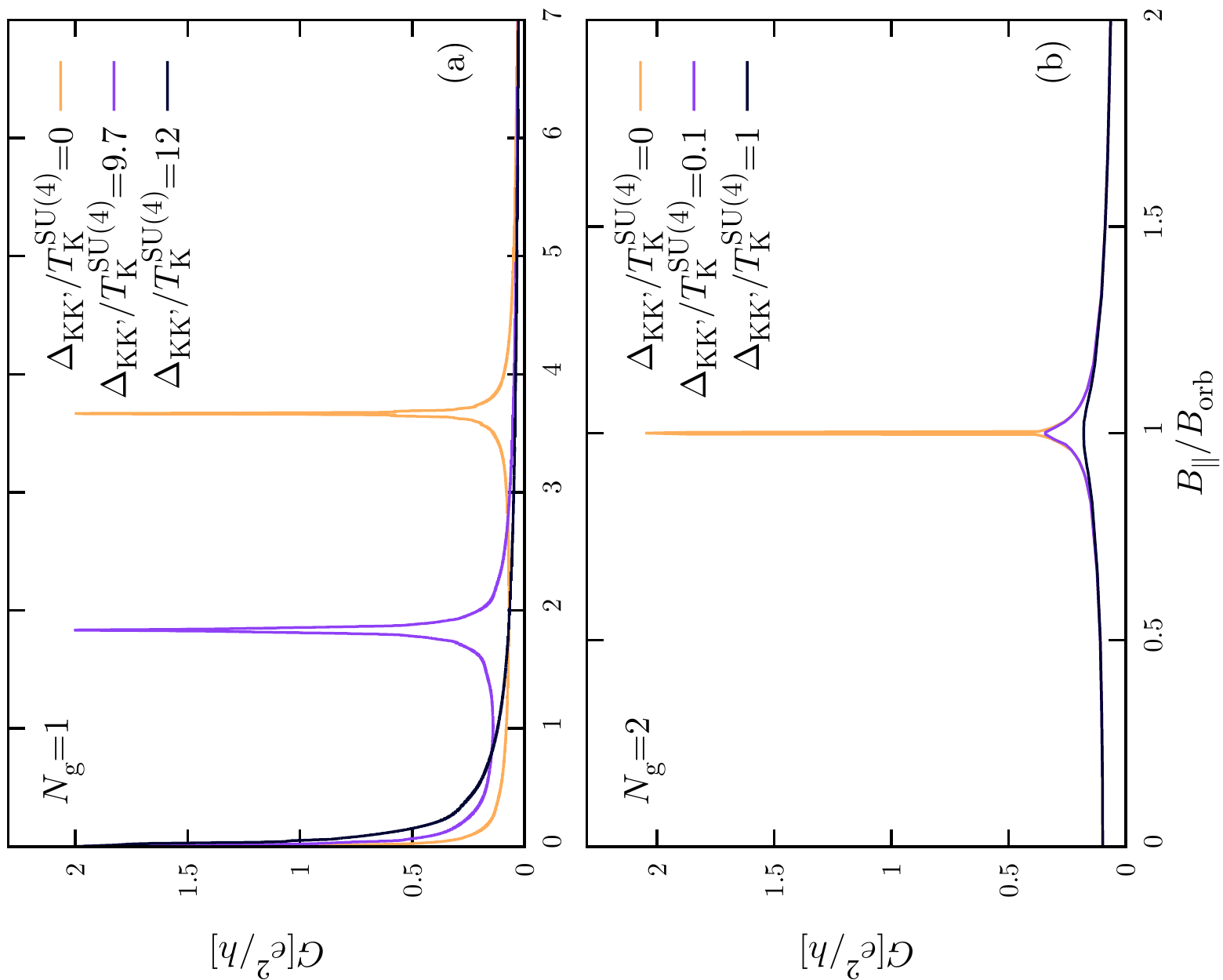}
  \caption{Linear conductance as a function of the parallel magnetic field (a) in the middle of the Coulomb valley,  $N_{\g}=1$, and (b) at the particle-hole symmetric point, $N_{\g}=2$ at $T=0$. $\Delta_{\SO}$ and $\Delta_{\KK}$ were varied such that $\Delta/T^{\SUf}_{\K}$ remained constant.}
\label{fig:cond_vs_BoverBorb}
\end{figure}

\paragraph{Linear Conductance} In \refig{fig:cond_vs_BoverBorb} we show the linear conductance of the system as a function of the applied parallel magnetic field and for a fixed ratio of $\Delta/T^{\SUf}_{\K}\simeq12$. In \refsubfig{fig:cond_vs_BoverBorb}{a} the conductance shows, as expected, the Kondo revival at $B_{\parallel}=B_{\s}$. Moreover, the width of the Kondo peak is proportional to $T_{\K}(\Delta)$. By increasing the valley mixing term $\Delta_{\KK}$ the resonance shifts towards $B_{\parallel}=0$ and, in case $\Delta_{\KK}/\Delta_{\SO}$ does not fulfill \refEq{eq:Bs_condition}, the Kondo revival disappears (black solid line). 

In the $N_{\g}=2$ valley, \refsubfig{fig:cond_vs_BoverBorb}{b}, the valley mixing term acts slightly differently. For $\Delta_{\KK}=0$ the $\SUf$-Kondo effect in $B_{\parallel}=0$ is essentially suppressed for this set of parameters ($\Delta/T^{\SUf}_{\K}\simeq12$) and hence a Kondo peak is present only at $B_{\parallel}=B_{\orb}$. Switching on the valley mixing, since $B_{\orb}$ is not $\Delta_{\KK}$ dependent, results in the suppression of this resonance without any shift. We notice that Kondo revivals have indeed been seen in experiments on CNT-dots in parallel fields~\cite{Jespersen2011}.

\subsection{Perpendicular magnetic field}
\label{sec:perpendicular_magn}

\begin{figure}[b!]
  \includegraphics[width=\columnwidth]
  {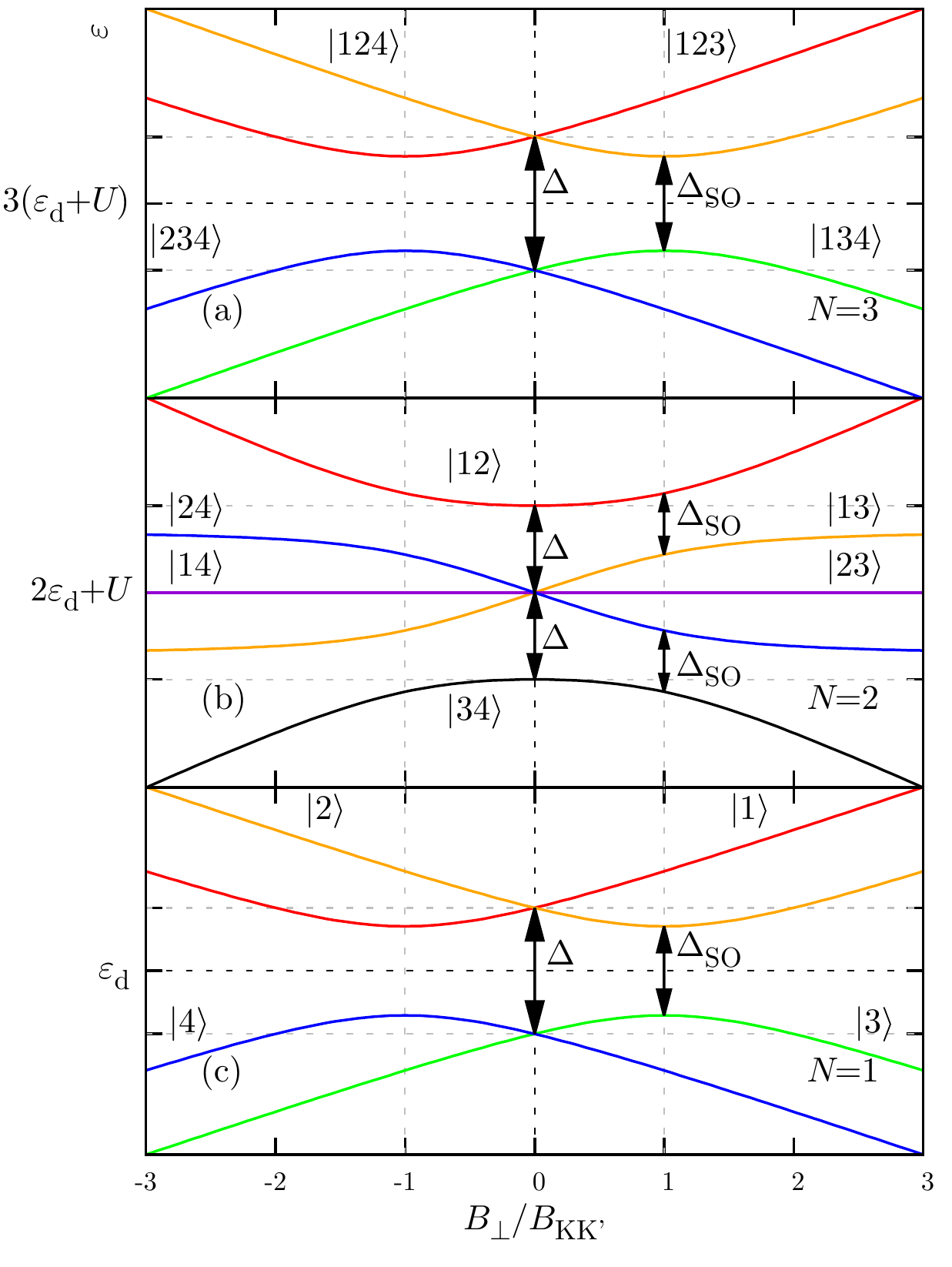}
  \caption{Perpendicular magnetic field dependence of the energy levels for (a) $N=3$, (b) $N=2$ and (c) $N=1$ particle number.}
\label{fig:energy_spectrum_perp}
\end{figure}

\paragraph{Spectrum} A magnetic filed perpendicular to the CNT axis couples only to the spin degree of freedom. As it follows from \reapp{subsec:specttrum_perp_field}, the energy levels of the  CNT Hamiltonian~\eqref{eq:H_dot} in the presence of a perpendicular magnetic field read
\begin{subequations}
\label{eq:energy_spectrum_perp}
 \begin{align}
     \varepsilon_{1,4}&=\varepsilon_{\dtxt}\pm\frac{1}{2}\sqrt{\Delta_{\SO}^2+\left(\Delta_{\KK}+g_{\s}B_{\perp}\right)^2},\\
     \varepsilon_{2,3}&=\varepsilon_{\dtxt}\pm\frac{1}{2}\sqrt{\Delta_{\SO}^2+\left(\Delta_{\KK}-g_{\s}B_{\perp}\right)^2},
  \end{align}
\end{subequations}
where $B_{\perp}$ is the amplitude of the perpendicular magnetic field.
The corresponding spectrum is shown in \refig{fig:energy_spectrum_perp}. It shows clear qualitative differences with respect to the parallel case. For example, no ground state crossing is observed for $N=1$ and $N=3$: the distance between the lowest energy levels first increases and then saturates to the valley mixing strength $\Delta_{\KK}$. 
An avoided crossing between excited states occurs for $B_{\perp}=\pm B_{\KK}$, where $B_{\KK}=\Delta_{\KK}/g_{\s}$. On the other hand, in the $N=2$ sector, the avoided crossing at $B_{\perp}=\pm B_{\KK}$ is between the ground state $\Ket{34}$ and the state $\Ket{24}$. Hence an exact crossing occurs for $\Delta_{\SO}=0$. Comparing \refigs{fig:energy_spectrum_perp}{fig:energy_spectrum} we indeed see that the SOI and the valley mixing exchange their role changing from parallel to perpendicular magnetic fields. Following the same argument of the previous section, this revels that the Kondo revival can be achieved only in the $N_{\g}=2$ valley for $B_{\perp}=B_{\KK}$. 

\begin{figure}[htbp!]
  \includegraphics[angle=-90,width=\columnwidth]
  {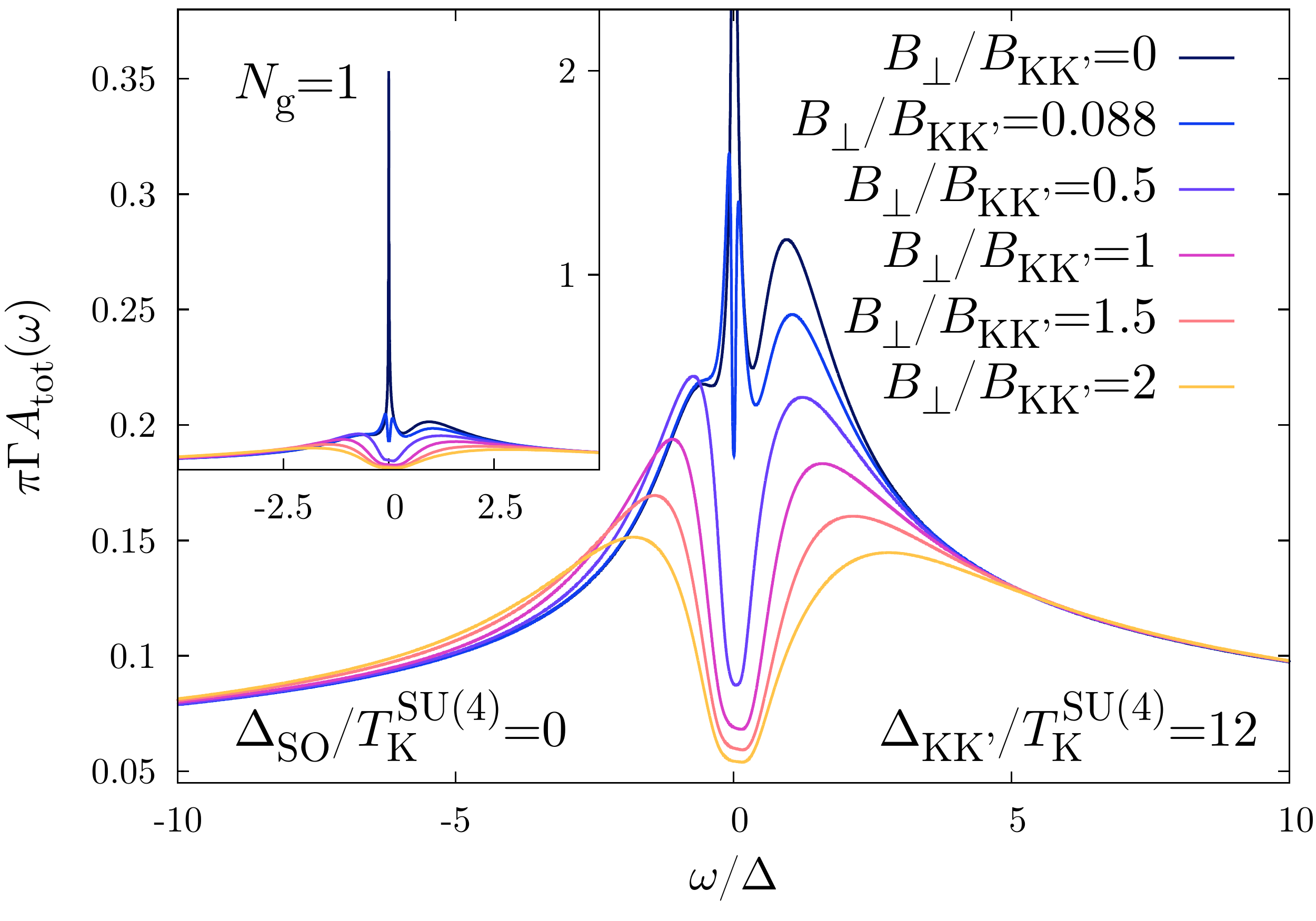}
\caption{Close-up structure of the total spectral function as a function of the frequency for several values of the perpendicular magnetic field, for $N_{\g}=1$ and $T=0$. Inset: Full spectral function, showing that the unitary limit is only obtained for $B_{\perp}=0$.}
\label{fig:sf_vs_omegaoverdelta_perp}
\end{figure}

\paragraph{Spectral function} We analyze the behavior of the total spectral function at $N_{\g}=1$ for several values of the perpendicular magnetic field in \refig{fig:sf_vs_omegaoverdelta_perp}. Since the Kondo revival is not possible in the first valley, the Kondo peak monotonically splits into two sub-peaks located at $\omega=\pm(\varepsilon_3(B_{\perp})-\varepsilon_4(B_{\perp}))$. Furthermore, two satellite peaks are visible. The rightmost resonance is the sum of two peaks in $A_1(\varepsilon)$ and $A_2(\varepsilon)$, whereas the leftmost one is essentially determined by the fourth component, $A_4(\varepsilon)$, corresponding to the ground state level. As in the parallel case, no splitting of the satellites is observed in the investigated magnetic field range. For $B_{\perp}\geq B_{\KK}/2$ the central Kondo peaks merge with the lateral satellites.
\begin{figure}[htbp!]
  \includegraphics[angle=-90,width=\columnwidth]
  {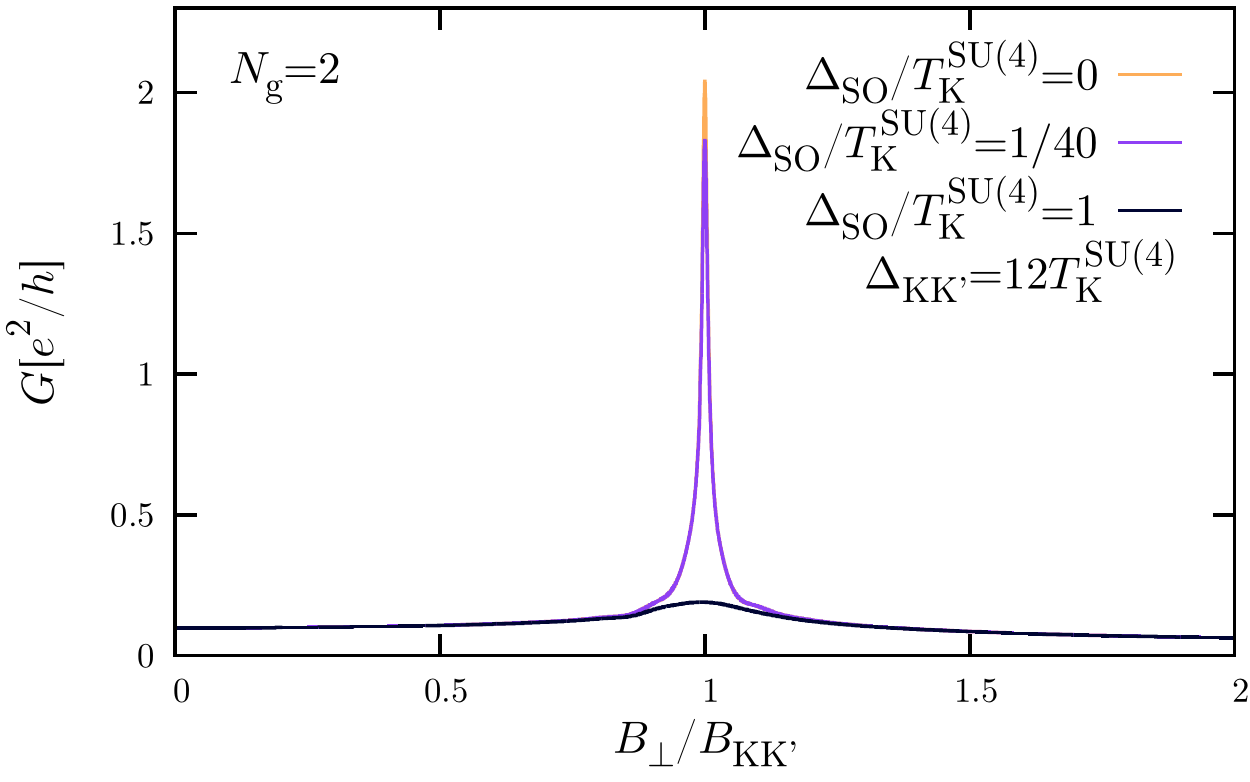}
  \caption{Linear conductance as a function of the perpendicular magnetic field at the particle-hole symmetric point, $N_{\g}=2$ at $T=0$.}
\label{fig:cond_vs_BoverBKK}
\end{figure}
\paragraph{Linear conductance} In \refig{fig:cond_vs_BoverBKK} we show the linear conductance as a function of the applied perpendicular magnetic field. A perfect Kondo revival is obtained only if $\Delta_{\SO}=0$; for $\Delta_{\SO}\neq0$ the Kondo resonance is suppressed because the degeneracy between the sates $\Ket{34}$ and  $\Ket{24}$ is lifted and the fluctuations of the quantum numbers referring to these states become energetically unfavorable.

\section{Conclusion}
\label{sec:conclusion}
In this work, 
we have investigated the transport  and spectral properties of  CNTs in the Kondo regime. To do that, we have constructed  a model that, for the sake of simplicity, assumes spin and valley quantum number conservation during tunneling processes, but  accounts for  the combined effects of spin-orbit interaction, valley mixing, and  electron-electron interaction, $U$, and finite magnetic field.

First, we carried out a detailed study of the cross-over between the $\SUf$ and $\SUt$-Kondo regime by   analyzing the dependence of the spectral functions and the linear conductance on the Kramers pair splitting, $\Delta$. We have shown by means of DM-NRG computations, in particular, that in the Kondo regime of a singly occupied CNT longitudinal mode a universal SU(4) conductance  is displayed for   $\Delta<T_{\K}^{\SUf}$, while for   $\Delta\gg T_{\K}^{\SUf}$ the conductance and the spectral functions  display characteristic SU(2) behavior with a strongly reduced Kondo temperature.    In the intermediate regime $T_{\K}^{\SUf}\ll \Delta\ll U$ we observed a $1/\Delta$ dependence of the Kondo temperature. This behavior can be explained in terms of simple scaling arguments, and is also in agreement with  exact Bethe Ansatz results obtained in case of infinitely strong  Coulomb repulsion~\cite{Sakano2006,Schlottmann1983}.

Finally,  we analyzed  the magnetic field dependence of the linear conductance and the spectral function, both in fields parallel and perpendicular to the CNT axis. Along the lines of \refcite{Galpin2010}, we derived the necessary conditions  to observe the Kondo revival, while extending our analysis to finite valley mixing values ($\Delta_{\KK}\neq0$) and to the case of perpendicular magnetic fields. We showed that in both parallel and perpendicular fields, the total spectral function of the system shows a four peak structure (two for the absorption and two for the emission processes). We showed that in the case of odd occupancy of a
CNT shell,  a large enough $\Delta_{\KK}$ can prevent the occurrence of a Kondo revival in parallel field; in perpendicular fields no revival is expected. Kondo revivals have indeed so far been observed experimentally only for parallel magnetic fields \cite{Jespersen2011,Schmid2015}.
In  case of a perpendicular magnetic field, the outer peaks can merge with those of the  split Kondo resonance, thereby leading to a two-peak structure (one for absorption and one for emission processes) for large magnetic field values compared to the $\SUf$-Kondo temperature  and $B_{\perp}>B_{\KK}/2$. 

\section*{Acknowledgement}
We gratefully thank Magdalena Marganska and Sergey Smirnov for the fruitful discussions. We acknowledge financial support through DFG Program No. GRK1570 and the Hungarian research grant No. OTKA K105149. 
G.Z. also thanks for its hospitality the Aspen Center for Physics, where part of this work has been completed.

\begin{appendix}

\section{Orthogonal transformations }
\label{app:ortho}

In this Appendix we present the construction of the Hamiltonian introduced in \refEq{eq:H_dot} in the Kramers basis  from an 
underlying Anderson model that takes into account the carbon nanotube structure~\cite{Marganska2014,Schmid2015}. 
In the absence of the valley mixing and spin-orbit interaction, the basis set $\{\Ket{K',\uparrow},\Ket{K',\downarrow},\Ket{K,\uparrow},\Ket{K,\downarrow}\}$, indexed by the valley and spin quantum numbers, is orthogonal. When $\Delta\ne 0 $ this is no longer true, and it is 
suitable to adopt instead the bonding (anti-bonding) representation $\left\{\Ket{a,\uparrow},\Ket{b,\uparrow},\Ket{a,\downarrow},\Ket{b,
\downarrow}\right\}$, which can be constructed as
\begin{subequations}
  \begin{align}
    \Ket{a,\sigma}&=\dop^{\dagger}_{+,\sigma}\Ket{0}=\frac{\Ket{K',\sigma}+\Ket{K,\sigma}}{\sqrt{2}},\\
    \Ket{b,\sigma}&=\dop^{\dagger}_{-,\sigma}\Ket{0}=\frac{\Ket{K',\sigma}-\Ket{K,\sigma}}{\sqrt{2}}.  
  \end{align}
\end{subequations}
In this basis the CNT Hamiltonian consists of several terms
\begin{align}
\label{eq:H_CNT}
  \Hop_{\CNT}&=\Hop^{(0)}_{\CNT}+\Hop_{e-e}+\Hop_{\B}\notag\\
  &=\Hop_{\dtxt}+\Hop_{\KK}+\Hop_{\SO}+\Hop_{e-e}+\Hop_{\B},
\end{align}
where $\Hop_{\dtxt}$ is the $\SUf$ invariant component and $\varepsilon_{\dtxt}$ the orbital energy which can be tuned through the applied gate voltage. Explicitly,
\begin{equation}
  \Hop_{\dtxt}=\varepsilon_{\dtxt}\sum_{i,\sigma=\pm}\dop^{\dagger}_{i,\sigma}\dop_{i,\sigma}\, .
\end{equation}
We add to the pure $\SUf$ term, respectively, the valley mixing (\refEq{eq:HKK}) and the SOI (\refEq{eq:HSO}) components:
\begin{subequations}
  \label{eq:dot_hamiltonian}
  \begin{align}
    &\Hop_{\KK}=\frac{\Delta_{\KK}}{2}\sum_{i,\sigma=\pm}i\dop^{\dagger}_{i,\sigma}\dop_{i,\sigma},\label{eq:HKK}\\
    &\Hop_{\SO}=\frac{\Delta_{\SO}}{2}\sum_{i,\sigma=\pm}\sigma\dop^{\dagger}_{-i,\sigma}\dop_{i,\sigma}.\label{eq:HSO}
   \end{align}
\end{subequations}
Notice that in the bonding/anti-bonding basis the valley mixing effect translates in an energy difference between the bonding and anti-bonding states  \cite{Marganska2014}. On the other hand, the SOI (due to curvature effects and relativistic correction to the CNT Hamiltonian) is off-diagonal.

The fourth term in ~\refEq{eq:H_CNT} describes the electron-electron interaction with $U$ associated to the charging energy of the dot
\begin{equation}
   \Hop_{e-e}=\frac{U}{2}\sum_{i\neq i_1=\pm}\sum_{\sigma,\sigma_1=\pm}\dop^{\dagger}_{i,\sigma}\dop_{i,\sigma}\dop^{\dagger}_{i_1,\sigma_1}\dop_{i_1,\sigma_1}.
\end{equation}
The external magnetic field enters through the last term in \refEq{eq:H_CNT}).

\subsection{Spectrum for zero magnetic field}
\label{subsec:spectrum_zero_B}

The single particle Hamiltonian~\refEq{eq:H_CNT} can be diagonalized by a unitary transformation, and the  new orthogonal basis is the Kramers basis
$\left\{\Ket{1},\Ket{4},\Ket{2},\Ket{3}\right\}$ used throughout of the present work. The unitary operator is
\begin{equation}
\label{eq:rotation_no_B}
	\mathcal{U}=
	\begin{pmatrix}
	    \cos\left(\theta\right)	&\sin\left(\theta\right)	&0	&0\\
	    -\sin\left(\theta\right)	&\cos\left(\theta\right)	&0	&0\\
	    0	&0	&\cos\left(\theta\right)	&-\sin\left(\theta\right)\\
	    0	&0	&\sin\left(\theta\right)	&\cos\left(\theta\right)
	\end{pmatrix},
\end{equation}
where the angle $\theta$ is given by $\tan(2\theta)=\Delta_{\SO}/\Delta_{\KK}$. In the diagonalized form, $\Hop_{\CNT}$ becomes the dot Hamiltonian  from~\refEq{eq:H_dot}.

We immediately notice that, because the two blocks in~\refEq{eq:rotation_no_B} yield the same eigenvalues, two pairs of degenerate doublets arise. These are the so-called Kramer pairs that, in our notation, are the couples of states $(1,2)$ and $(3,4)$. As such, the states within each Kramers pair are related through the time reversal operator $\Top$ as it is sketched in \refig{fig:CPT}. Additionally, valley reversal, governed by the anti-unitary operator $\Pop$, relates the couples $(1,4)$ and $(2,3)$ originating from the two sub-blocks. In the Kramers basis $\Top$ and $\Pop$ are given by
\begin{subequations}
\label{eq:CPT}
  \begin{align}
    \Top&=\kappa\left(\dop^{\dagger}_{2}\dop_{1}+\dop^{\dagger}_{4}\dop_{3}-\hc\right),\\
    \Pop&=\kappa\left(\dop^{\dagger}_{4}\dop_{1}-\dop^{\dagger}_{3}\dop_{2}-\hc\right),
  \end{align}
\end{subequations}
where $\kappa$ stands for the complex conjugation. Finally, defining $\Cop= \Pop\cdot\Top^{-1}$, it is possible to relate the couples $(1,3)$ and $(2,4)$ to each other through the unitary operator
\begin{equation}
      \Cop=\dop^{\dagger}_1\dop_3+\dop^{\dagger}_2\dop_4+\hc
\end{equation}

\subsection{Spectrum for finite parallel magnetic field}
\label{subsec:specttrum_par_field}

An external magnetic field parallel to the CNT axis couples to both the spin degree of freedom and the ``orbital'' one. Thus, in the bonding(anti-bonding) basis, $\Hop_{\B}$ reads \cite{Schmid2015}
\begin{align}
    \Hop_{\B}&=\Hop^{\parallel}_{\B-\s}+\Hop^{\parallel}_{\B-\orb}\notag\\
    &=\frac{1}{2}g_{\s}B_{\parallel}\sum_{i,\sigma=\pm}\sigma\dop^{\dagger}_{i,\sigma}\dop_{i,\sigma}+g_{\orb}B_{\parallel}\sum_{i,\sigma=\pm}\dop^{\dagger}_{-i,\sigma}\dop_{i,\sigma},
\end{align}
where $B_{\parallel}$ is the amplitude of the parallel magnetic field. Thus, the diagonalized single particle Hamiltonian is
\begin{equation}
 \mathcal{U}^{\dagger}\left(B_{\parallel}\right)\left(\Hop^{(0)}_{\CNT}+\Hop_{\B}\right)\mathcal{U}\left(B_{\parallel}\right)=\sum^{4}_{j=1}\varepsilon_{j}\dop^{\dagger}_{j}\dop_{j}\\
\end{equation}
where 
\begin{equation*}
\label{eq:rotation_B_parallel}
	\mathcal{U}\left(B_{\parallel}\right)=
	\begin{pmatrix}
	    \cos\left(\theta^+\right)	&\sin\left(\theta^+\right)	&0	&0\\
	    -\sin\left(\theta^+\right)	&\cos\left(\theta^+\right)	&0	&0\\
	    0	&0	&\cos\left(\theta^-\right)	&-\sin\left(\theta^-\right)\\
	    0	&0	&\sin\left(\theta^-\right)	&\cos\left(\theta^-\right)
	\end{pmatrix}.
\end{equation*}
Here we defined $\theta^{\pm}$ in such a way that $\tan(2\theta^{\pm})=(\Delta_{\SO}\pm2g_{\orb}B_{\parallel})/\Delta_{\KK}$. Notice that $\mathcal{U}\left(B_{\parallel}\right)$ posses a similar block structure as in~\refEq{eq:rotation_no_B}. However, due to $\theta^{+}\neq\theta^{-}$, time reversal symmetry is broken.

\subsection{Spectrum for finite perpendicular magnetic field}
\label{subsec:specttrum_perp_field}
An external magnetic field perpendicular to the CNT axis couples, differently to the parallel case, only to the spin degree of freedom \cite{Marganska2014,Schmid2015}. Its action reads
\begin{equation}
    \Hop_{\B}=\frac{1}{2}g_{\s}B_{\perp}\sum_{i,\sigma=\pm}\dop^{\dagger}_{i,\bar{\sigma}}\dop_{i,\sigma}.
\end{equation}
The transformation that diagonalizes the single particle Hamiltonian can be decomposed as a product of two orthogonal matrices:
\begin{align*}
    \mathcal{U}\left(B_{\perp}\right)&=\mathcal{O}_1\left(B_{\perp}\right)\mathcal{O}_2\left(B_{\perp}\right),\\
    \mathcal{O}_1\left(B_{\perp}\right)&=
	\begin{pmatrix}
	    \cos\left(\theta^+\right)	&\sin\left(\theta^+\right)	&0	&0\\
	    -\sin\left(\theta^+\right)	&\cos\left(\theta^+\right)	&0	&0\\
	    0	&0	&\cos\left(\theta^-\right)	&\sin\left(\theta^-\right)\\
	    0	&0	&-\sin\left(\theta^-\right)	&\cos\left(\theta^-\right)
	\end{pmatrix},\\
    \mathcal{O}_2\left(B_{\perp}\right)&=\frac{1}{\sqrt{2}}
	\begin{pmatrix}
	  1	&0	&1	&0\\
	  0	&1	&0	&-1\\
	  1	&0	&-1	&0\\
	  0	&1	&0	&1
	\end{pmatrix},
\end{align*}
where $\theta^{\pm}$ is such that $\tan(2\theta^{\pm})=\Delta_{\SO}/(\Delta_{\KK}\pm g_{\s}B_{\perp})$. We notice that $\mathcal{U}\left(B_{\perp}\right)$ mixes both the spin and the bonding(anti-bonding) degrees of freedom, whereas, in the previous case, $\mathcal{U}\left(B_{\parallel}\right)$ was mixing only the latter one.

\section{Universality and the fixed points}
\label{app:fermi_liquid}
\begin{figure}[!ht]
  \includegraphics[width=\columnwidth]
  {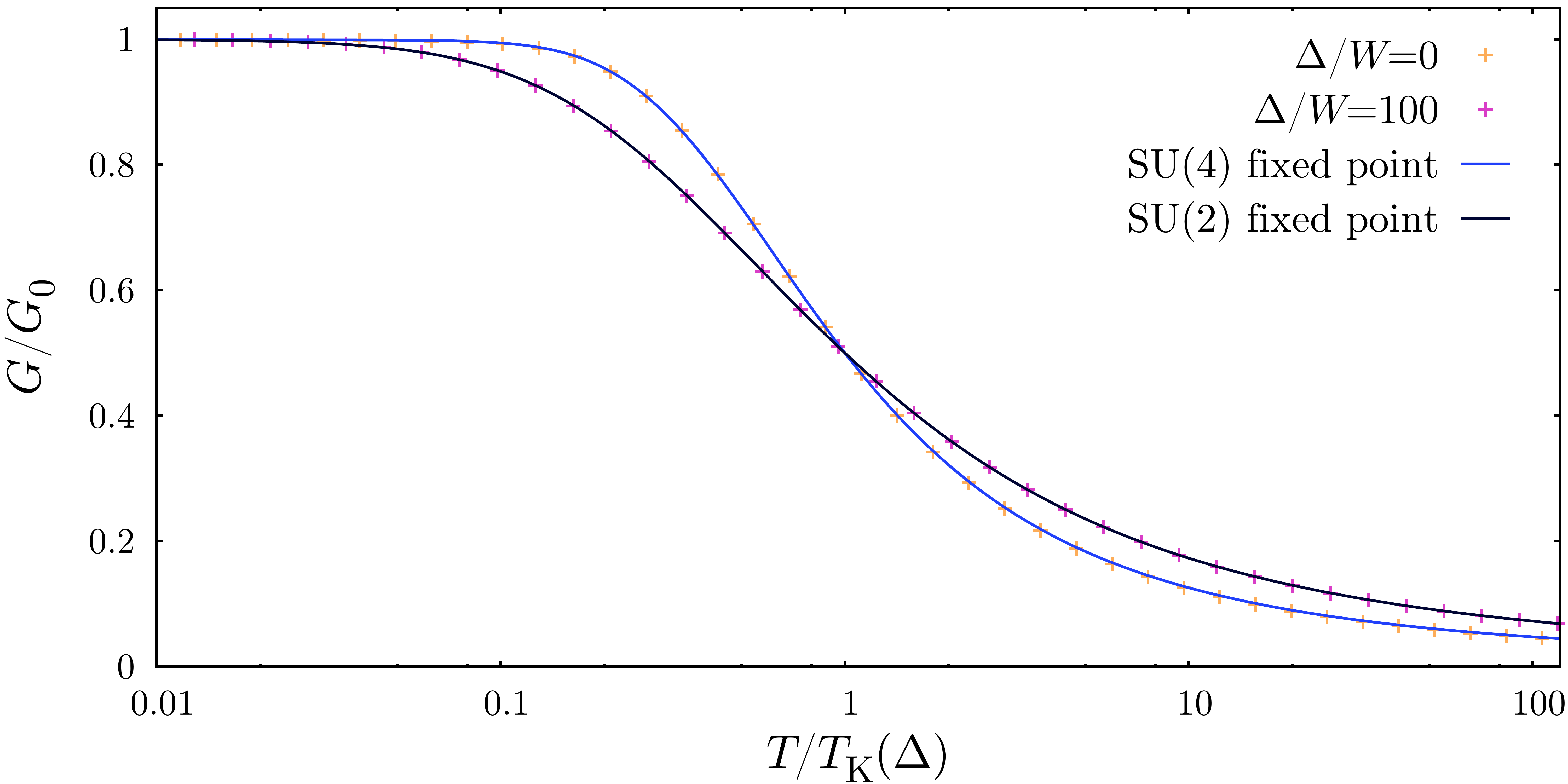}
  \caption{Comparison between the temperature dependence of the conductance in the limits when $\Delta\to 0$ and $\Delta\to \infty$ (symbols) and the SU(2) and SU(4) universal curves for the conductance (solid lines).}
\label{fig:fixed_points}
\end{figure}
We devote this Appendix to a brief analysis of the properties of the $\SUt$ and $\SUf$ limiting cases in the $N_{\g}=1$ valley. 
When $\Delta=0$ we expect the system to be at the $\SUf$ fixed point. In  \refig{fig:fixed_points} we compare on one side, the 
conductance as computed with the help of \refEq{eq:G} for $\Delta=0$, with the universal curve for $G (T/T_{\K})$ as obtained from an
underlying Kondo model with SU(4) symmetry, and we find a perfect agreement. Increasing $\Delta$, the system flows away from the SU(4) and 
towards the SU(2) fixed point. As the other curve in \refig{fig:fixed_points} shows, for large enough $\Delta$'s, i.e. $\Delta\gg W$, the system has already reached the SU(2) fixed point.  

\begin{figure}[b!]
  \includegraphics[width=\columnwidth]
  {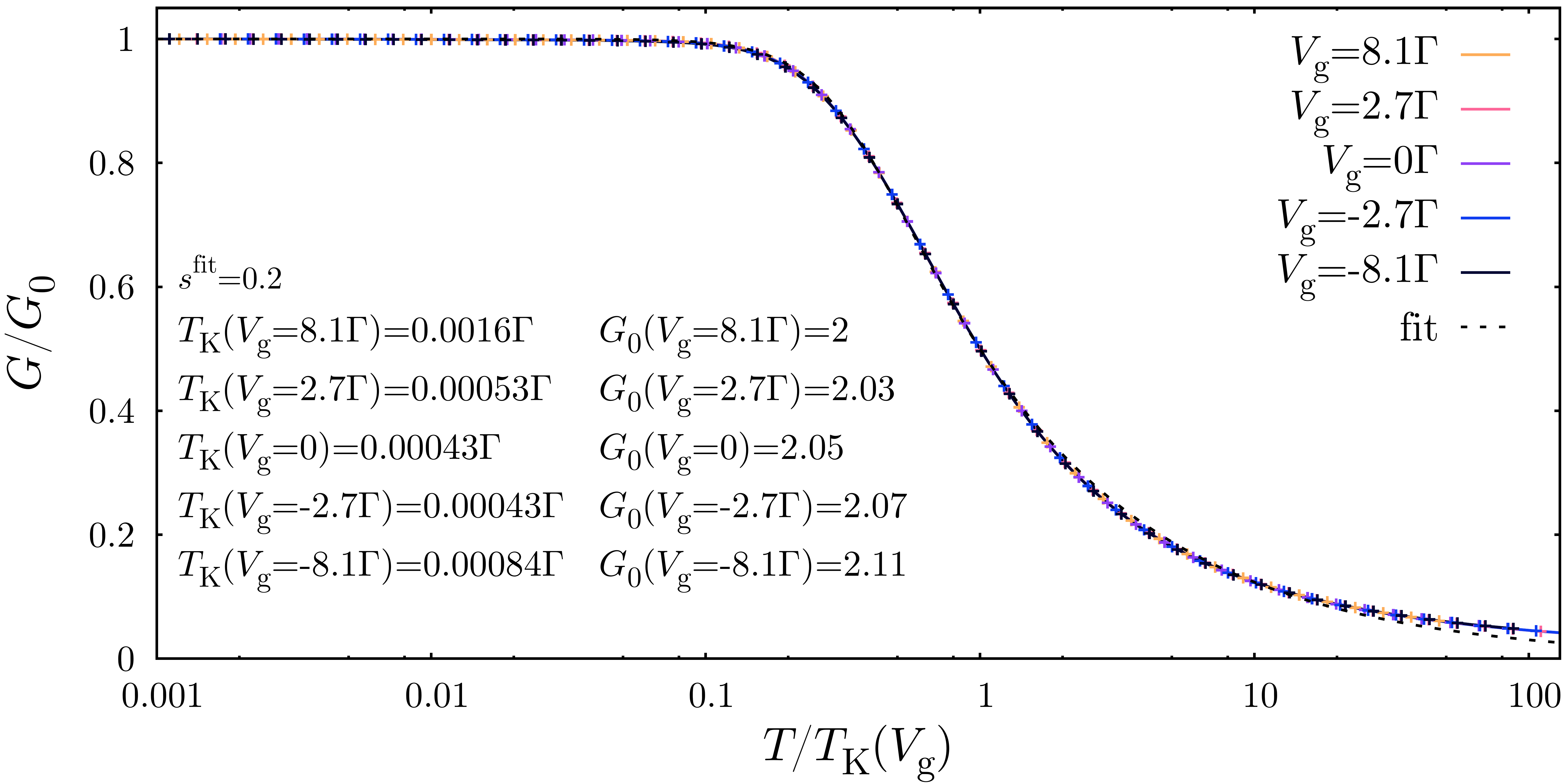}
  \caption{Universal scaling of the conductance as a function of $T/T_{\K}$ for different gate voltages and comparison with against the heuristical curve from~\refEq{eq:fit_curve} with $n=3$. The other parameters are: $U=W$, $\Gamma=U/50$, $T=0$, $\varepsilon_{\dtxt}=-U/2$ and $\Delta=0$. Here $G_{0}$ is the unitary conductance in unit $e^2/h$.}
\label{fig:G_vs_TTK_5vgs}
\end{figure}

Recently, in  \refcite{Keller2014} a heuristical analytical expression was proposed for the  universal
conductance, of the form
\begin{equation}
 \label{eq:fit_curve}
 \frac{G\left(T/T_{\K}\right)}{G_{0}} =\left(1+\left(2^{1/s}-1\right)\cdot\left(\frac{T}{T_{\K}}\right)^{n}\right)^{-s},
\end{equation}
in order to reproduce the leading-order in the temperature expansion predicted by the conformal field theory \cite{Hur2007,Mora2009} for the $\SUf$-Kondo Hamiltonian. In this case the leading order is predicted to be cubic,
$G(T/T_{\K})/G_0\simeq 1-\alpha\cdot (T/T_{\K})^3+\dots$, with $\alpha$ a constant of the order $\sim 1$,  even 
though the system has still a Fermi liquid character. This sets $n=3$ in \refEq{eq:fit_curve} in contrast to the case  $n=2$ for the $\SUt$ 
symmetry.

In \refig{fig:G_vs_TTK_5vgs} we compare the conductance with this heuristic curve for different gate voltages. 
From the fits of the DM-NRG data (\refig{fig:G_vs_TTK_5vgs}), in the range $0\leq T/T_{\K}\leq 1$, we obtained a good estimate for $s=0.202\pm0.002$ in agreement with \refcite{Keller2014}. 
The heuristic curve reproduces very well the DM-NRG results in a wide range  of temperatures, $0\leq T/T_{\K}\le 10$, and deviations become visible only for large enough temperatures. 

Trying to fit the data for $\Delta\gg W$ in \refig{fig:fixed_points} with $n=2$ we found $s=0.215\pm0.002$, in agreement with the predictions
from the conformal field theory.

\section{Numerical renormalization group approach}
\label{app:nrg}

We solve the Hamiltonian~\eqref{eq:H_tot} using the numerical renormalization group approach. The core of the NRG is a logarithmic
discretization of the conduction band with a parameter $\Lambda\simeq 2$, followed by a mapping of the Hamiltonian to a semi-infinite chain. 
In this way the problem can be solved perturbatively, as the hopping couplings along the chain, $t_{n}^{j}\sim \Lambda^{-n/2} $ decrease exponentially~\cite{Bulla2008}. The conduction band Hamiltonian becomes
\begin{equation}
 \Hop_{\chain}=\sum_{j=1}^4\sum^{+\infty}_{n=0}t^{j}_n\,\fop^{\dagger}_{j,n}\,\fop_{j,n+1}+\hc.
\end{equation}
Here  $\fop^{\dagger}_{j,n}$ are the fermionic creation operators at the $n$-th site in the $j$-th channel. The impurity is sitting at
site -1 and is coupled to the site $n=0$
\begin{equation}
 \Hop_{\tun}=V\sum_{j=1}^4\fop^{\dagger}_{j,0}\dop_{j}+\hc
\end{equation}
As the dot Hamiltonian~\eqref{eq:H_dot} is not modified by this procedure, the total Hamiltonian consists now of four spinless conduction bands coupled to a complex impurity composed of the dot degrees of freedom. 

Along the NRG procedure it is crucial to use the symmetries of the system in order to achieve numerically reliable results. 
When $\Delta=0$, the model is SU(4) invariant, as the total Hamiltonian commutes with the SU(4)-spin operator
\begin{equation}
  \label{eq:su4_generators}
  \hat{\vec{J}}^{\SUf}=\frac{1}{2}\sum^{+\infty}_{n=-1}\sum_{j,j'=1}^4\fop^{\dagger}_{j,n}\vlambda_{jj'}\fop_{j',n},
\end{equation}
where $\vlambda=(\lambda_1,\dots,\lambda_{15})$ is a set of matrices defining the  generators for the $\SUf$ algebra (generalized Gell Mann matrices~\cite{Sbaih2013} for example).
As discussed in  \resec{subsec:symmetries}, there are also two $\Uone$ symmetries corresponding to the conservation of charge in each of the Kramers channels. 
In the NRG language the  generators are
\begin{equation}
\label{eq:charge_op}
  \Qop_{\kappa}=\frac{1}{2}\sum_{j\in\kappa}\sum^{+\infty}_{n=-1}\left ( \fop^{\dagger}_{j,n}\,\fop_{j,n}-{\frac{1}{2}}\right).\\
\end{equation}
 
Since the generators for spin and for the charges commute among themselves,
the system has a global $\Uone\otimes \Uone\otimes\SUf$ symmetry. 
The spin orbit and valley mixing perturbations, i.e. $\Delta\ne 0$, break the global symmetry down to the $\Uone\otimes\Uone\otimes\SUt\otimes\SUt$, 
generated by the charge and the SU(2)-spin operators 
\begin{equation}
  \label{eq:su2_generators_nrg}
  \hat{\vec{J}}_{\kappa}=\frac{1}{2}\sum^{+\infty}_{n=-1}\sum_{j,j'\in \kappa}\fop^{\dagger}_{j,n}\vsigma_{jj'}\fop_{j',n},
\end{equation}
acting on the two Kramers doublets. Here $\vsigma=(\sigma_x,\sigma_y,\sigma_z)$ is the vector of Pauli matrices.

\end{appendix}


\end{document}